\documentclass[a4paper,11pt]{article}
\pdfoutput=1 
\usepackage{jheppub} 
\usepackage{bbm,bm,graphicx,mathtools,color,hyperref,slashed}
\usepackage{amssymb,amsmath}
\usepackage[T1]{fontenc}

\usepackage[all]{xy}

\graphicspath{{./Figures/}}

\newcommand{\diff}{\mathrm{d}}
\newcommand{\p}{\partial}
\newcommand{\ve}{\varepsilon}
\newcommand{\Diff}{{\mathcal{D}}}

\newcommand{\be}{\begin{equation}}      
\newcommand{\ee}{\end{equation}}      
\newcommand{\bea}{\begin{eqnarray}}      
\newcommand{\eea}{\end{eqnarray}}

\newcommand{\tr}{\mathrm{tr}}

\newcommand{\im}{\mathrm{i}}

\newcommand{\rme}{\mathrm{e}}

\newcommand{\eq} {equation}

\newcommand{\NN} {\nonumber}

\preprint{YITP-20-99}

\title{Topological aspects of $4$D Abelian lattice gauge theories with the $\theta$ parameter}

\author{Masazumi Honda, }

\emailAdd{masazumi.honda~AT~yukawa.kyoto-u.ac.jp}

\author{Yuya Tanizaki}

\affiliation{Yukawa Institute for Theoretical Physics, Kyoto University, Kyoto 606-8502, Japan}

\emailAdd{yuya.tanizaki~AT~yukawa.kyoto-u.ac.jp}

\abstract{
We study a four-dimensional $U(1)$ gauge theory with the $\theta$ angle, which was originally proposed by Cardy and Rabinovici. 
It is known that the model has the rich phase diagram 
thanks to the presence of  both electrically and magnetically charged particles.
We discuss the topological nature of the oblique confinement phase of the model at $\theta=\pi$, 
and show how its appearance can be consistent with the anomaly constraint. 
We also construct the $SL(2,\mathbb{Z})$ self-dual theory out of the Cardy-Rabinovici model by gauging a part of its one-form symmetry. 
This self-duality has a mixed 't~Hooft anomaly with gravity, and its implications on the phase diagram is uncovered. 
As the model shares the same global symmetry and 't~Hooft anomaly with those of $SU(N)$ Yang-Mills theory, studying its topological aspects would provide us more hints to explore possible dynamics of non-Abelian gauge theories with nonzero $\theta$ angles. 
}

\begin{document}
\maketitle
\section{Introduction}
\label{sec:introduction}
Quark confinement is a basic feature of the strong interaction, while it is an important and unsolved problem to derive it in any analytical methods based on non-Abelian Yang-Mills (YM) theories. 
One of the famous scenarios of quark confinement is the dual superconductivity, which assumes that the YM vacuum is caused by condensation of monopoles~\cite{Nambu:1974zg, Mandelstam:1974pi, tHooft:1977nqb, Polyakov:1975rs}. 
When both electric and magnetic particles exist, 
we can expect rich structure of possible phases, such as the Coulomb phase, Higgs phase, confinement phase, etc. 
The monopole is quite useful to understand the qualitative effect of the $\theta$ angle and more exotic phases can appear for $\theta\not=0$~\cite{tHooft:1981bkw}, as the magnetic monopole acquires the fractional electric charge proportional to $\theta$ and it is known as the Witten effect~\cite{Witten:1979ey}. 

Cardy and Rabinovici proposed 
a simple model in order to study these nontrivial dynamics quantitatively in the presence of $\theta$~\cite{Cardy:1981qy, Cardy:1981fd}, and we call it the Cardy-Rabinovici model in this paper. 
It is the lattice $U(1)$ gauge theory on the four-dimensional cubic lattice coupled to the charge-$N$ Higgs particles, and the monopoles can be described by the violation of Bianchi identity at the lattice scale. 
In the continuum formulation, we introduce the $\theta$ angle as a coupling to the instanton density, but such topologies are suffered from lattice discretization and lose some of important features valid in the continuum (See Refs.~\cite{Sulejmanpasic:2019ytl, Gattringer:2018dlw} for recent developments on topologies of lattice $U(1)$ gauge theories).
In the Cardy-Rabinovici model, the $\theta$ angle is introduced to reproduce the Witten effect.  
Cardy and Rabinovici have conjectured that the phase diagram has a rich structure in the space of the coupling $g^2$ and the $\theta$ angle by applying the heuristic free-energy argument to identify the condensate in the vacua~\cite{Cardy:1981qy}.

An interesting property of this model is the appearance of oblique confinement phase, originally proposed by 't~Hooft~\cite{tHooft:1981bkw}. 
When $\theta\simeq \pi$, the monopole costs not only magnetic energies but also electric energies due to the Witten effect, which suggests that its condensation becomes less probable in the strong coupling regime around $\theta\simeq\pi$. 
Instead, a bound state of two monopoles and one Higgs particle does not carry net electric charge because the electric charge induced by the Witten effect is canceled by that of the Higgs particle, and the bound state is more likely to condense near $\theta\simeq \pi$.

There is a recent progress on the YM dynamics 
at $\theta=\pi$ by the new anomaly matching condition~\cite{Gaiotto:2017yup}.  
The four dimensional pure $SU(N)$ YM theory enjoys the $\mathbb{Z}_N$ one-form symmetry, $\mathbb{Z}_N^{[1]}$. 
The one-form symmetry is not an ordinary symmetry in the sense that it does not transform local operators \cite{Gaiotto:2014kfa}.  
Instead, the one-form symmetry acts on line operators, which describe the world-line of test particles, 
or very heavy quark in the YM theory.
In the $SU(N)$ YM theory, 
we can measure $N$-ality of their electric charges so we have $\mathbb{Z}_N$ symmetry acting on the Wilson loops.  
Like ordinary symmetries, 
we can consider the gauging operation of $\mathbb{Z}_N^{[1]}$, but we should introduce two-form gauge fields.  
In Ref.~\cite{Gaiotto:2017yup}, it is found that the $\mathsf{CP}$ symmetry at $\theta=\pi$ is violated under the presence of 
nontrivial two-form background gauge fields for even $N$, 
and this anomaly is renormalization-group invariant. 
When $N$ is odd, more subtle condition, called global inconsistency, gives the constraint on the phase diagram, instead~\cite{Gaiotto:2017yup, Tanizaki:2017bam, Kikuchi:2017pcp, Tanizaki:2018xto, Komargodski:2017dmc, Karasik:2019bxn, Cordova:2019jnf, Cordova:2019uob}. 
The $\mathsf{CP}$ symmetry at $\theta=\pi$ has been suspected to be spontaneously broken~\cite{Dashen:1970et, Witten:1980sp, DiVecchia:1980yfw, Ohta:1981ai, Creutz:1995wf, Smilga:1998dh, Witten:1998uka, DiVecchia:2017xpu}, and this argument unveils that it is partly required by kinematical reasoning.  

In this paper, we 
show that the Cardy-Rabinovici model has the same structure of the symmetry and anomaly
as the $SU(N)$ YM theory. 
As all the dynamical electric particles have the $U(1)$ gauge charge $N$, 
the model enjoys the $\mathbb{Z}_N^{[1]}$ symmetry, 
and we introduce the $\theta$ angle in a way that it becomes $2\pi$ periodic, $\theta\sim \theta+2\pi$, in the low-energy limit. 
As a result, the theory at $\theta=\pi$ acquires the $\mathsf{CP}$ symmetry, 
and it has the mixed anomaly with $\mathbb{Z}_N^{[1]}$ 
as in the case of $SU(N)$ YM theory.
The presence of the 't~Hooft anomaly tells us more details on the phase diagram, and we confirm the consistency between the anomaly constraint and the proposed phase diagram by clarifying the topological aspects of each phase. 
In particular, we study topological aspects of oblique confinement, which crucially depend on whether $N$ is even or odd. 
When $N$ is even, the oblique confinement phase is a $\mathbb{Z}_2$ topological order at low energies with the spontaneous symmetry breaking,
\be
\mathbb{Z}_N^{[1]}\to \mathbb{Z}_{N/2}^{[1]}. 
\ee
When $N$ is odd, the oblique confinement phase is trivial as intrinsic topological orders, while it should be separated from the usual confinement phase as they are different symmetry-protected topological (SPT) states with $\mathbb{Z}_N^{[1]}$ to match the global inconsistency. 

There is another interesting aspect of the Cardy-Rabinovici model. 
In Ref.~\cite{Cardy:1981fd},  the effective Lagrangian of charges and monopoles is obtained by integrating out the fluctuation of photons, and the effective Lagrangian has the $SL(2,\mathbb{Z})$ self-duality. 
Indeed, we find that the local dynamics is identical under the $SL(2,\mathbb{Z})$ transformation, so the local quantities, such as the free-energy density, must be the same under the duality transformation. 
However, we show that the $SL(2,\mathbb{Z})$ duality does not extend to the global aspects of the theory. 
In some cases, the duality operation exchanges the topologically trivial and nontrivial phases, and we clarify that this is because the duality transformation does not preserve the one-form symmetry\footnote{
These structures are quite parallel to the $\mathcal{N}=4$ super-Yang Mills theory as discussed in Ref.~\cite{Aharony:2013hda}.
}.

We therefore construct the model, whose local dynamics is the same with that of the Cardy-Rabinovici model, while the $SL(2.\mathbb{Z})$ self-duality extends to the global aspect of the theory. 
It is obtained by considering the Cardy-Rabinovici model with the charge $N=M^2$, 
and then we gauge the subgroup $\mathbb{Z}_M^{[1]}$ of the one-form symmetry $\mathbb{Z}_N^{[1]}=\mathbb{Z}_{M^2}^{[1]}$. 
This gauged Cardy-Rabinovici model enjoys the $\mathbb{Z}_M^{[1]}\times \mathbb{Z}_M^{[1]}$ symmetry, and the first factor denotes an electric one-form symmetry and the another does a magnetic one. 
It turns out that the gapped phases of the gauged Cardy-Rabinovici model always show the $\mathbb{Z}_M$ topological order, and this is because of the mixed anomaly between electric and magnetic $\mathbb{Z}_M^{[1]}$ symmetries. 

As the electric and magnetic one-form symmetries are isomorphic, we can show that the gauged model enjoys the $SL(2,\mathbb{Z})$ duality. 
In this model, however, the $\theta$ angle is no longer $2\pi$ periodic, $\theta\not\sim \theta+2\pi$, and the map $\theta\to \theta+2\pi$ must be identified as one of the generators of $SL(2,\mathbb{Z})$. 
This $SL(2,\mathbb{Z})$ duality is the same with that of the pure Maxwell theory. 
The partition function of the Maxwell theory is not $SL(2,\mathbb{Z})$ invariant on general spin four-manifolds, but instead transforms as the modular form~\cite{Witten:1995gf, Verlinde:1995mz}. 
We can regard this as a mixed anomaly between $SL(2,\mathbb{Z})$ duality and the Lorentz invariance~\cite{Seiberg:2018ntt}, so we find another anomaly constraint on the phase diagram.

Organization of this paper is as follows. 
In Sec.~\ref{sec:review}, we give a review on the Cardy-Rabinovici model. 
We will see that the model is expected to have a rich phase structure based on the heuristic discussion on the free energy of  world lines of dyonic excitations. 
We will also give a review on the $SL(2,\mathbb{Z})$ self-duality about the local dynamics, emphasizing that it does not necessarily extend to the global nature of the model. 
In Sec.~\ref{sec:anomaly_CRmodel}, we study the topological aspects of the Cardy-Rabinovici model, which partly justified the conjectured structure of the phase diagram.  
The anomaly matching plays an important role for this purpose, and we give the formal continuum definition of the Cardy-Rabinovici model in order to compute its anomaly in a clear manner. 
Using the continuum reformulation, we obtain the mixed anomaly, or global inconsistency, for $\mathbb{Z}_N^{[1]}$ and $\mathsf{CP}$ at $\theta=\pi$. 
In Sec.~\ref{sec:gaugedCRmodel}, we give the gauged Cardy-Rabinovici model using the continuum formulation, and study the $SL(2,\mathbb{Z})$ self-duality as a genuine property of the theory. 
We discuss the mixed anomaly between $SL(2,\mathbb{Z})$ and gravity to constrain the possible phase diagram. 
In Sec.~\ref{sec:summary}, we summarize the results and discuss possible implications to the non-Abelian YM dynamics with nonzero $\theta$ angles. 

\section{Review on Cardy-Rabinovici lattice gauge model with the $\theta$ angle}\label{sec:review}
In this section, we give a brief review of the work by Cardy and Ravinobici on the lattice $U(1)$ gauge theory with the $\theta$ angle~\cite{Cardy:1981qy, Cardy:1981fd}. 
This model is expected to show the rich phase structure due to the various types of charge, monopole, and dyon condensations~\cite{Cardy:1981qy}. 
Moreover, the local dynamics of this model enjoys the $SL(2,\mathbb{Z})$ self-duality, which constrains possible structures of the phase diagram~\cite{Cardy:1981fd}. 

\subsection{Description of the Cardy-Rabinovici model}
The four-dimensional lattice gauge theory in Refs.~\cite{Cardy:1981qy, Cardy:1981fd} is defined as follows, and we call it the Cardy-Rabinovici model. 
The spacetime is assumed to be the four-torus $T^4$, and it is discretized as the square lattice. 
Instead of the usual formulation of lattice gauge theory,
here we use a formulation known as the Villain form of the lattice $U(1)$ gauge theory~\cite{Villain:1974ir}
to conveniently describe the $\theta$ angle and magnetic degrees of freedom.
In the Villain form,
noting the structure $U(1)=\mathbb{R}/\mathbb{Z}$,
the $U(1)$ gauge field $a$ on the discretized torus is introduced 
as a pair $(\widetilde{a}_\mu, s_{\mu\nu})$, 
where $\widetilde{a}_\mu$ is the $\mathbb{R}$-valued link variable
and $s_{\mu\nu}$ is the $\mathbb{Z}$-valued plaquette variable. 

First the kinetic term of the $U(1)$ gauge field is given by
\be
S_{\rm kin} =\frac{1}{2g^2} \sum_{(x,\mu,\nu)} f_{\mu\nu}(x)^2 .
\ee
where $f=\diff a$ is the field strength,
\be
f_{\mu\nu}=\p_\mu \widetilde{a}_\nu - \p_\nu \widetilde{a}_\mu-2\pi s_{\mu\nu} .
\ee
Note that the field strength 
is invariant under the $\mathbb{R}$-valued $0$-form gauge transformation 
\begin{\eq}
\widetilde{a}_\mu\to \widetilde{a}_\mu+\p_\mu \lambda^{(0)} ,
\end{\eq}
and the $\mathbb{Z}$-valued $1$-form gauge transformation
\begin{\eq}
\widetilde{a}_\mu \to \widetilde{a}_\mu+2\pi \lambda^{(1)}_\mu, \quad
s_{\mu\nu}\to s_{\mu\nu}+\p_\mu \lambda^{(1)}_\nu-\p_\nu \lambda^{(1)}_\mu .
\label{eq:Z_1form}
\end{\eq}
This lattice discretization of $U(1)$ gauge theory allows us to define the monopole current, 
\be
m_\mu(\tilde{x})= \frac{1}{2} \ve_{\nu \mu \lambda \sigma} \p_\nu s_{\lambda \sigma}(x). 
\ee
Here, $\tilde{x}$ is the site on the dual lattice, $\tilde{x}=x+\frac{1}{2}(\hat{1}+\hat{2}+\hat{3}+\hat{4})$, and thus $m_\mu(\tilde{x})$ is the $\mathbb{Z}$-valued link variable on the dual lattice. 
By definition, it satisfies the conservation law,
\be
\p_\mu m_\mu=0,
\ee
and thus the configuration of $m_\mu$ can be regarded as the closed world-line of magnetically charged particles. 

Next we describe the matter part.
The electric matter field in Refs.~\cite{Cardy:1981qy, Cardy:1981fd} is introduced as the closed world-line representation. 
It is defined as the $\mathbb{Z}$-valued link variable $n_\mu$, satisfying the constraint,
\be
\p_\mu n_\mu = 0.
\label{eq:electric_conservation}
\ee
When the electric matter has the charge $N$
so that the theory enjoys $\mathbb{Z}_N$ one-form symmetry,
its minimal coupling term in the Lagrangian is given by
\be
\im N  n_\mu (x) \widetilde{a}_\mu (x). 
\ee
The gauge invariance of this coupling is ensured by the conservation law (\ref{eq:electric_conservation}). 
Indeed, summing up $n_\mu$ restricts $\tilde{a}_\mu$ into $\frac{2\pi}{N}\mathbb{Z}$, so the theory becomes the lattice $\mathbb{Z}_N$ gauge theory.

\begin{figure}[t]
\centering
\includegraphics[scale=0.67]{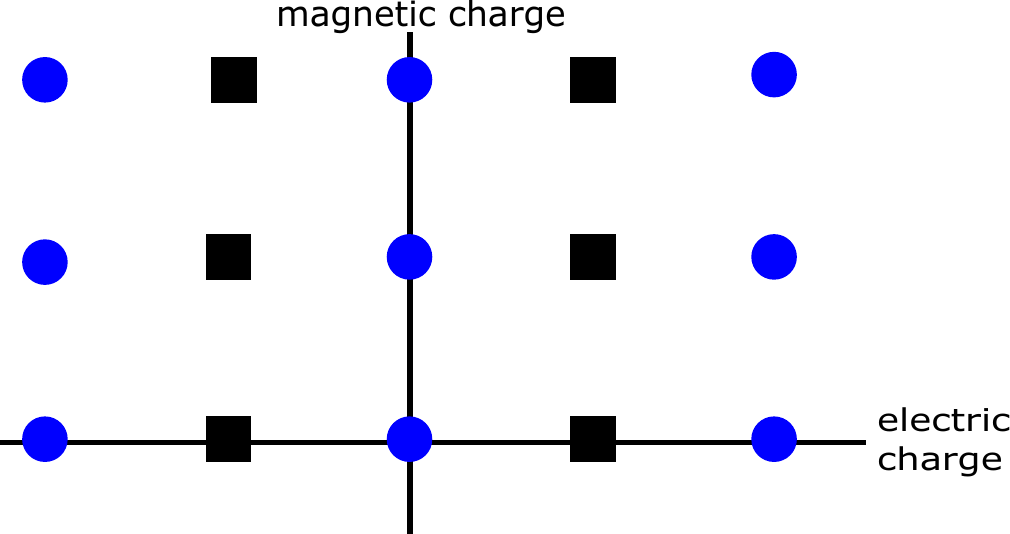}\qquad 
\includegraphics[scale=0.67]{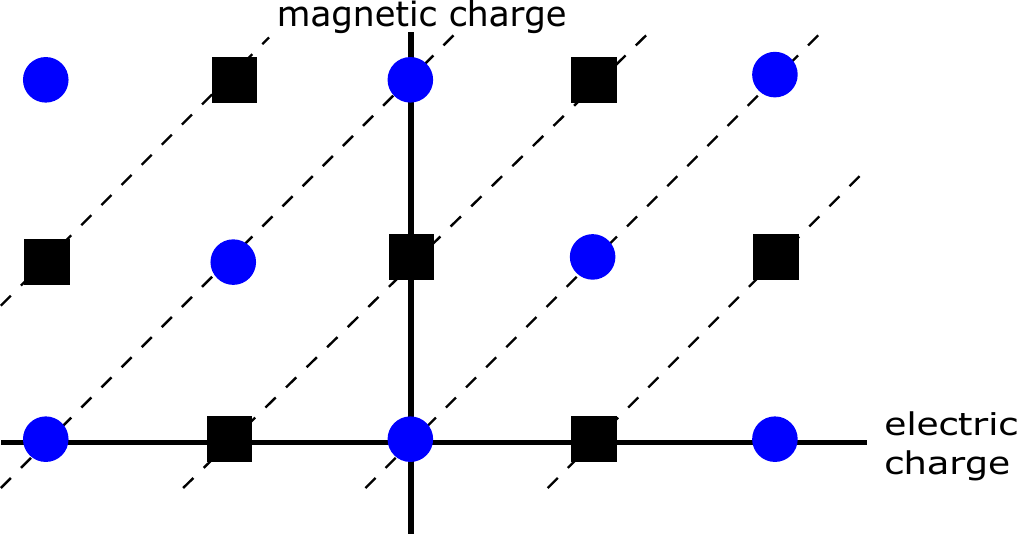}
\caption{
Charge lattices at $\theta=0$ (left) and $\theta=\pi$ (right) for $N=2$. 
The blue blobs show the charges of dynamical excitations, 
while the black squares show those of test particles satisfying the Dirac quantization condition. 
The electric charges of dynamical excitations are quantized to the multiples of $N$, $N\mathbb{Z}$, and the other charges can be present only as test particles. 
}
\label{fig:charge_lattice}
\end{figure}

Now, let us introduce the $\theta$ parameter to this lattice model. 
This can be done by noticing that the magnetic monopole acquires the electric charge $\frac{\theta}{2\pi}$ by the Witten effect~\cite{Witten:1979ey}. 
In order to reproduce this nature, we replace $n_\mu$ by 
\be
\widetilde{n}_\mu(x)=n_\mu(x)+\frac{\theta}{2\pi} \sum_{\tilde{x}} F(x-\tilde{x}) m_\mu(\tilde{x}). 
\ee
Here, $F(x-\tilde{x})$ is a short-ranged function in order to relate the dual lattice and the original lattice. 
Although the choice of $F$ is arbitrary, the details of $F$ is expected not to affect the low-energy dynamics of this model. Now, the Lagrangian of the matter fields becomes
\be
S_{\rm mat}
=\im N  \sum_{(x,\mu)} \tilde{n}_\mu (x) \widetilde{a}_\mu (x) .
\ee
At sufficiently low energies, the distinction between the original and dual lattices is expected to be no longer important\footnote{In this paper, we assume that the Cardy-Rabinovici model has a nice continuum limit or UV completion. }, and then we may simply write the effective electric current as 
\be
\widetilde{n}_\mu(x) = n_\mu(x) + \frac{\theta}{2\pi} m_\mu(x). 
\ee
Since both $n_\mu$ and $m_\mu$ are $\mathbb{Z}$-valued link variables, there is an emergent $2\pi$ periodicity of the $\theta$ parameter at low energies, because $\tilde{n}_\mu$ is invariant under $\theta\to \theta+2\pi $ and $n_\mu \to n_\mu-m_\mu$. 
As an example,
we draw the list of charged particle excitations in Fig.~\ref{fig:charge_lattice} for $N=2$ at $\theta=0$ and $\pi$. 
Since the gauge group is $U(1)$,
all the points in the charge lattice allow test particles, 
which can be introduced as the genuine line operators.
The charges of the dynamical excitations, denoted with the blue blobs,
are more restricted to preserve the $\mathbb{Z}_N$ one-form symmetry.

Combining these data, 
the action of the model is
\bea
S [\tilde{a}_\mu, s_{\mu\nu}, n_\mu] 
&=&S_{\rm kin} [\tilde{a}_\mu, s_{\mu\nu} ] +S_{\rm mat} [\tilde{a}_\mu, s_{\mu\nu}, n_\mu] \NN\\
&=&\frac{1}{2g^2}\sum_{(x,\mu,\nu)} f_{\mu\nu}(x)^2
+\im N \sum_{(x,\mu)}\left(n_\mu(x)+\frac{\theta}{2\pi} \sum_{\tilde{x}} F(x-\tilde{x}) m_\mu(\tilde{x})\right)\widetilde{a}_\mu(x). \NN\\
\label{eq:Lagrangian1} 
\eea
Then the partition function of the model is defined by
\be
Z
=\mathrm{Tr} \left[ \rme^{-S[\tilde{a}_\mu, s_{\mu\nu}, n_\mu] }\right],
\ee
where the symbol ``$\mathrm{Tr}$'' stands for integrations over the $\mathbb{R}$-valued link variable $\tilde{a}_\mu$,
and summations over the $\mathbb{Z}$-valued plaquette variable $s_{\mu\nu}$ 
and $\mathbb{Z}$-valued link variable $n_\mu$ satisfying (\ref{eq:electric_conservation}).
Since the Lagrangian (\ref{eq:Lagrangian1}) is quadratic in $\widetilde{a}_\mu$, we can integrate it out exactly. 
The effective Lagrangian can be written only in terms of $n_\mu$ and $m_\mu$~\cite{Cardy:1981qy}: 
\bea
S_{\mathrm{eff}}
&=&\frac{2\pi^2}{g^2}\sum_{\tilde{x}, \tilde{x}'} m_\mu(\tilde{x}) G(\tilde{x}-\tilde{x}') m_\mu(\tilde{x}') +\frac{N^2 g^2}{2}\sum_{x,x'}\widetilde{n}_\mu(x) G(x-x')\widetilde{n}_\mu(x')\nonumber\\
&&- \, \im N \sum_{\tilde{x},x}m_\mu(\tilde{x}) n_\nu(x) \Theta_{\mu\nu}(\tilde{x}-x). 
\label{eq:Lagrangian2}
\eea
Here, $G(x-x')$ is the lattice massless Green function, and $\Theta_{\mu\nu}$ is an angle-valued function defined in Ref.~\cite{Cardy:1981qy}, which expresses the angle between the Dirac string emitted from the monopole $m_\mu(\tilde{x})$ and the four-vector $(\tilde{x}-x)$. 
It is $\Theta_{\mu\nu}$ that ensures the Dirac-Schwinger-Zwanziger quantization condition~\cite{Dirac:1931kp, Schwinger:1966nj, Zwanziger:1968rs}, as it jumps by $2\pi$ when an electric charge goes through the world-sheet of the Dirac string. Therefore, the Dirac string is not physically observable only if the charge quantization is satisfied. 

\subsection{Phase diagram via the free-energy argument}
\label{sec:free_energy}
In order to get physical intuitions on the Cardy-Rabinovici model \eqref{eq:Lagrangian1},
let us study the phase structure 
based on a simple free-energy argument in Ref.~\cite{Cardy:1981qy}
(see also Refs.~\cite{Banks:1977cc, Savit:1977fw}).  
As we will see soon,
the argument is heuristic and 
details on quantitative structures should not be taken seriously.
The steps are as follows:
\begin{enumerate}
\item Given the parameters $(g,\theta )$,
we estimate the internal energy of a particle excitation with fixed electric and magnetic charges, 
which forms a closed world line in the four dimensional spacetime.

\item We judge that the particle can condense 
if the energy is smaller than the entropy of the loop. 

\item When there are several candidates for charged particles to condense, 
we pick up the minimal energy one. 
On the other hand, when none of the charges can condense,
we interpret that the realized phase is the Coulomb phase.

\end{enumerate}

The electric and magnetic charges of dynamical excitations are labeled by integers $(n,m)\in \mathbb{Z}\times \mathbb{Z}$ as 
\be
\left(N\Bigl(n+\frac{\theta}{2\pi}m\Bigr), m\right) .
\label{eq:label_charges}
\ee
Then we estimate the energy of the loop excitation by extracting the short-range part of the self Coulomb interaction from (\ref{eq:Lagrangian2}), 
\be
\left( \frac{N^2 g^2}{2}\left(n+\frac{\theta}{2\pi}m\right)^2 + \frac{2\pi^2}{g^2}m^2\right) G(0)L ,
\ee
where
the long-range part is neglected because it can be screened by the presence of other loops. 
Noting that
the entropy of loops with the length $L$ in the $d$-dimensional cubic lattice is roughly given by $L \ln(2d-1)$, 
the particle $(n,m)$ can condense if~\cite{Cardy:1981qy} 
\be
\ve_{n,m}(g,\theta)
\equiv \left( \frac{N g^2}{2\pi }\left(n+\frac{\theta}{2\pi}m\right)^2 + \frac{2\pi}{N g^2}m^2\right)N < C,
\label{eq:ansatz_free_energy}
\ee
where $C={\ln 7/\pi G(0)}$, while the value of the constant $C$ should not be taken too seriously.
This expression suggests that the condensation becomes harder
as $N$ takes larger values, 
so there is more chance for the Coulomb phase for larger $N$.  
We note that the lattice Monte Carlo simulation of this model is possible when $\theta=0$~\cite{Creutz:1979kf, Creutz:1979zg}, and the result is consistent with the free-energy argument with roughly $C\sim 6$. 

\begin{figure}[t]
\centering
\includegraphics[scale=0.5]{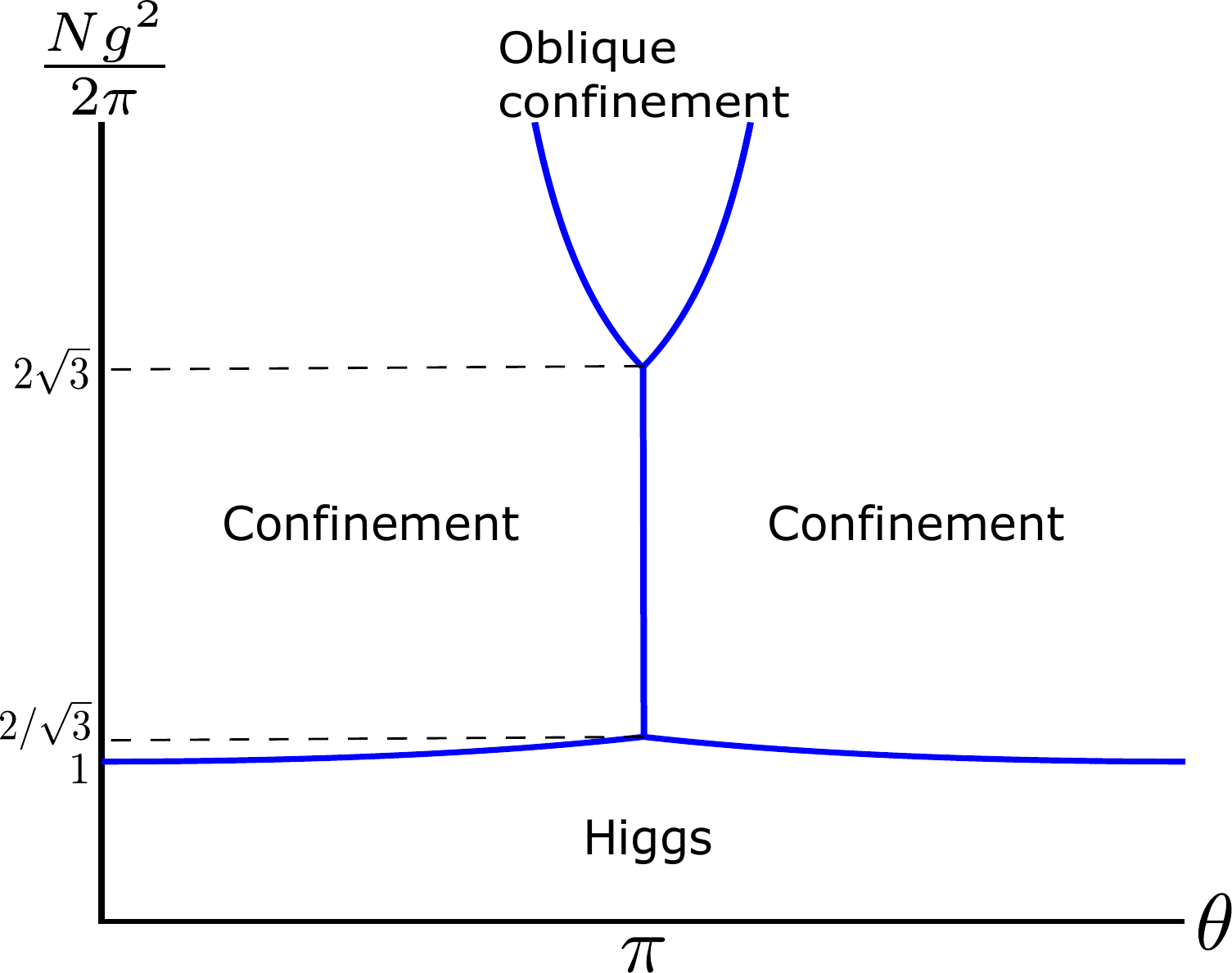}
\includegraphics[scale=0.5]{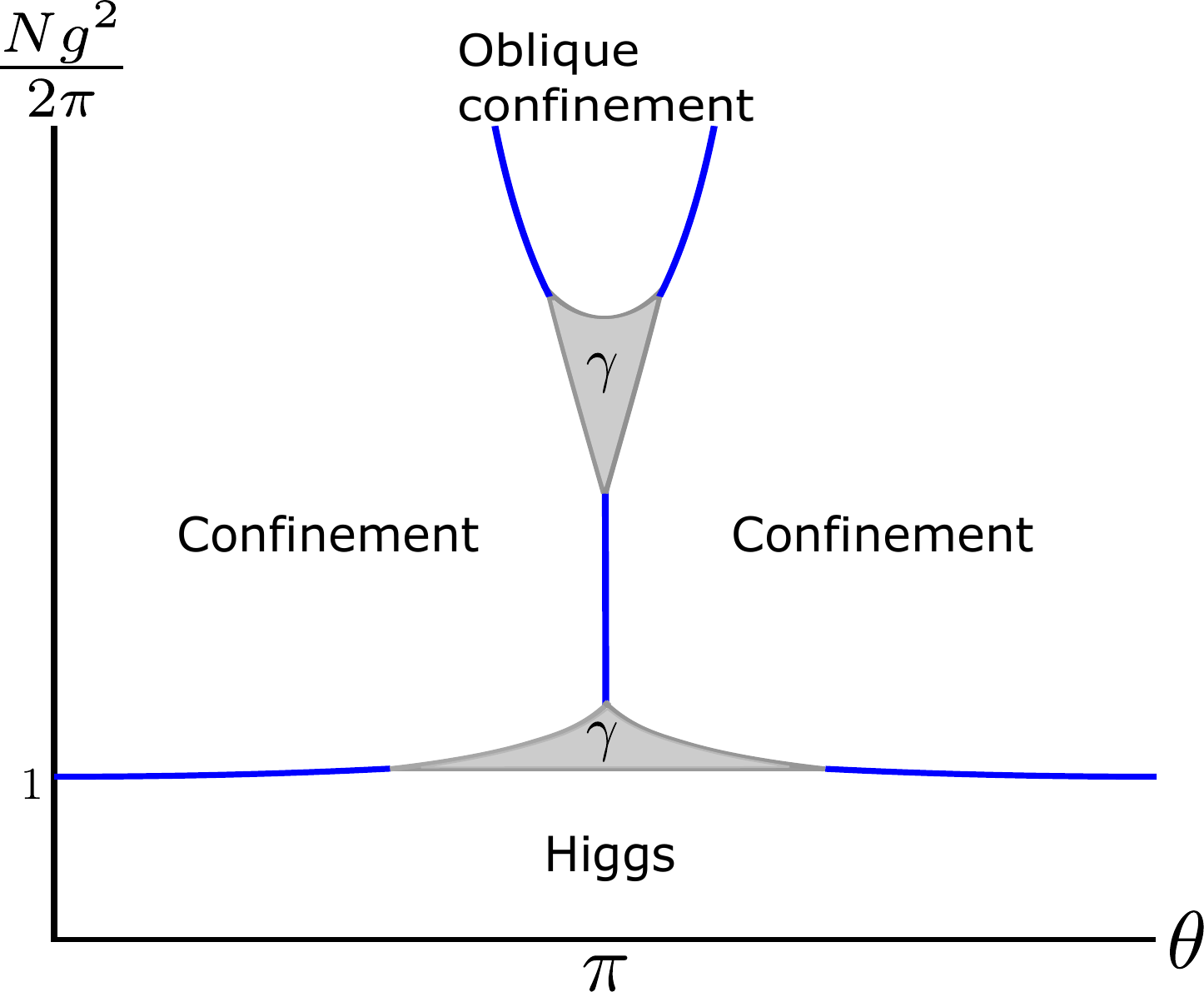}
\caption{
Phase diagrams for $0<\theta<2\pi$ and $N g^2/2\pi<5$ when $C/N>2/\sqrt{3}$ (left) and $C/N<2/\sqrt{3}$ (right) based on the simple free energy argument. 
In the right figure, the Coulomb phase appears in the gray regions
denoted by $\gamma$. 
The weak coupling region is governed by the Higgs phase, while the strong coupling region by the confinement phase. 
For $0<\theta<\pi$, the confinement is caused by the condensation of the monopole, $(n,m)=(0,1)$, while for $\pi<\theta<2\pi$, it is caused by the condensation of the dyon, $(n,m)=(-1,1)$. 
For sufficiently strong coupling, $N g^2/2\pi>2\sqrt{3}$, the oblique confinement mode appears near $\theta=\pi$, which is the condensation of $(n,m)=(-1,2)$. 
Blue solid curves separate these gapped phases by the phase transitions. 
}
\label{fig:phase_diagram1}
\end{figure}

When $C/N>2/\sqrt{3}\simeq 1.15$, it turns out that 
there always exists a particle to condensate and 
therefore the Coulomb phase does not appear in the phase diagram. 
Thus the system is always in gapped phases for $C/N>2/\sqrt{3}$ at any $(g,\theta )$.
The phase diagram is given in Fig.~\ref{fig:phase_diagram1}, and the left and right figures show the cases for $C/N> 2/\sqrt{3}$ and $C/N< 2/\sqrt{3}$, respectively. 
When $C/N<2/\sqrt{3}$, the Coulomb phase appears in the gray regions, which is denoted by $\gamma$. 
Other phases are the gapped phases due to the condensation of charged particles. 

\subsubsection*{Higgs phase}
The Higgs phase is defined by the condensation of the electric charge, given by $(n,m)=(1,0)$, so its energy is 
\be
E_{\mathrm{Higgs}}(g,\theta)=\ve_{1,0}(g,\theta)=\frac{Ng^2}{2\pi}N. 
\ee
This does not have any $\theta$ dependence. 
Another important point is that it does not have the $1/g^2$ term as the Higgs particle does not carry the magnetic charge. Therefore, in the weak-coupling regime, the Higgs mode should be the most favored gapped phase.

\subsubsection*{Confinement phase}
Confinement is caused by the condensation of magnetically charged particles. The most naive one is the condensation of the magnetic monopole, $(n,m)=(0,1)$, and its energy is 
\be
\ve_{0,1}(g,\theta )
=\left(\frac{2\pi}{N g^2}+\frac{N g^2}{2\pi} \left( \frac{\theta}{2\pi}\right)^2\right)N. 
\ee
The internal energy increases quadratically as a function of $\theta$. 

Since the $\theta$ angle has the periodicity $2\pi$ in the continuum limit, the above expression means that the monopole condensation should not be a valid description for large values of $\theta$. 
Therefore, we should take into account the possibility of the dyon condensation, too~\cite{tHooft:1981bkw, Witten:1980sp}. 
When the dyon with the charge $(n,1)$ condenses, its free energy becomes 
\be
\ve_{n,1}(g,\theta )
=\left( \frac{2\pi}{N g^2} +\frac{N g^2}{2\pi} \left( \frac{\theta}{2\pi}+n\right)^2\right)N. 
\ee
Among these states, we select the minimal energy state as the confinement phase at $\theta$, so the energy density is 
\be
E_{\mathrm{confined}}(g,\theta )
=\min_{n\in \mathbb{Z}} \left(\ve_{n,1}(g,\theta )\right)
= \min_{n\in \mathbb{Z}} \left( \frac{2\pi}{N g^2} +\frac{N g^2}{2\pi} \left( \frac{\theta}{2\pi}+n\right)^2\right)N. 
\ee
If we take $\theta =(2n_\theta -1)\pi +\delta\theta$
with $n_\theta \in\mathbb{Z}$ and $0\leq \delta \theta <2\pi$,
then it is solved as
\be
E_{\mathrm{confined}}(g,\theta )
= \left. \ve_{n,1}(g,\theta ) \right|_{n=-n_\theta}
=\left( \frac{2\pi}{N g^2} +\frac{N g^2}{2\pi} \left( \frac{\delta\theta}{2\pi} -\frac{1}{2} \right)^2\right)N ,
\ee
where the solution is unique for $\delta \theta \neq 0$    
while there is another solution $n=-n_\theta +1$ for $\delta \theta = 0$  with the same minimum.
For example, when $-\pi<\theta<\pi$, the monopole $(n,m)=(0,1)$ would condense, but when $\pi<\theta<3\pi$, the dyon $(n,m)=(-1,1)$ would condense, and there is the phase transition at $\theta=\pi$.

\subsubsection*{Oblique confinement phase}
Oblique confinement, which was originally proposed by 't~Hooft~\cite{tHooft:1981bkw}, is characterized by the condensation of the higher monopole charge, such as $(n,m)=(-1,2)$. 
It is the new phase with the condensation of magnetically charged particles, and it turns out to be energetically favored only if the charge lattice is oblique due to the Witten effect. 

The charge, $(-1,2)$, can be regarded as the composite particle of monopole $(0,1)$ and dyon $(-1,1)$, and its condensation energy is given by 
\be
\ve_{-1,2}(g,\theta )
=\left(4 \cdot \frac{2\pi}{N g^2} +\frac{N g^2}{2\pi} \left( \frac{\theta}{\pi}-1\right)^2\right)N. 
\ee
Near $\theta=\pi$, this condensation does not cost the electric energy at all, while the both Higgs and confinement phases cost some of them  (see the right panel of  Fig.~\ref{fig:charge_lattice}). 
Therefore, if the electric coupling $g^2$ is sufficiently large, the oblique confinement can overcome the usual confinement, and it indeed appears when $\frac{Ng^2}{2\pi}>2\sqrt{3}$ at $\theta=\pi$ in Fig.~\ref{fig:phase_diagram1}. 

As $\ve_{-1,2}(\theta)$ is not $2\pi$ periodic in $\theta$, we again consider the list of charges $(2n'-1,2)$ and pick up the minimal energy one:
\be
\min_{n'\in \mathbb{Z}} \left( \ve_{2n'-1,2}(g,\theta )\right) 
=\min_{n'\in\mathbb{Z}}\left(4 \cdot \frac{2\pi}{N g^2} +\frac{N g^2}{2\pi} \left( \frac{\theta+2\pi n'}{\pi}-1\right)^2\right)N.
\ee
This is the free-energy density for the oblique confinement phase with manifest $2\pi$ periodicity. 
When $\theta\simeq \pi$, the state $n'=0$ is chosen, but, for example, when $\theta\simeq -\pi$, the different state $n'=1$ is selected. 

According to this free-energy argument, it becomes evident that more exotic oblique confinement phase appears~\cite{Cardy:1981qy}. 
When $\theta/2\pi$ is a rational number, i.e. $\theta/2\pi=-p/q$, 
the condensate of the charge $(p,q)$ does not cost any electric energies 
and thus it is preferred at sufficiently strong couplings. 
In order to discuss those oblique confinement phases in a systematic manner, it is convenient to resort to self-dual nature of the Cardy-Rabinovici model~\cite{Cardy:1981fd}, as we shall review in the next subsection. 

\subsection{$SL(2,\mathbb{Z})$ duality of the free-energy density and the phase diagram}
\label{sec:review_duality}
So far, we draw the phase diagram, Fig.~\ref{fig:phase_diagram1}, just by putting an ansatz (\ref{eq:ansatz_free_energy}) for the condensation of the particle type $(n,m)$. 
The structure of the phase diagram is claimed to be justified by using the self-dual nature of the model~\cite{Cardy:1981fd} (see also Refs.~\cite{Elitzur:1979uv, Horn:1979fy} for the model without $\theta$). 
The following discussion is true for the local dynamics such as the free-energy density, but, as we shall see in the later sections, it does not always generalize to other observables. 

Let us introduce the complex coupling,
\be
\tau=\frac{\theta}{2\pi}+ \im \frac{2\pi}{N g^2}, 
\ee
which is in the upper half plane. 
Assuming that the distinction between the original and dual lattices disappears at the low energies, the Lagrangian (\ref{eq:Lagrangian2}) is invariant under the $SL(2,\mathbb{Z})$ duality transformation
\begin{\eq}
\tau \to \frac{a\tau +b}{c\tau +d} \quad (a,b,c,d\in\mathbb{Z}\ \  {\rm and}\ \ ad-bc =1) ,
\end{\eq}
which is generated by the $S$ and $T$ transformations~\cite{Cardy:1981fd}:
\be
S: \tau\to -\frac{1}{\tau},\quad (n,m)\to (-m,n),
\label{eq:S_transformation}
\ee
and 
\be
T:\tau\to \tau+1,\quad (n,m)\to (n-m,m). 
\label{eq:T_transformation}
\ee
We can express the self-duality group as 
\be
SL(2,\mathbb{Z})=\Bigl\langle S,T \, \Bigl | \, S^2=(ST^{-1})^3\, , \, S^4=1 \Bigr\rangle. 
\ee
We note that $S^2=\mathsf{C}$ trivially acts on the space of coupling 
but it flips both electric and magnetic fields,
and therefore it is identified as the charge conjugation symmetry
\begin{\eq}
\mathsf{C}:\tau\to \tau, \quad (n,m)\to (-n,-m) . 
\end{\eq}
There is also the $\mathsf{CP}$ transformation,
\be
\mathsf{CP}:\tau \to -\overline{\tau},\quad  (n,m)\to (-n,m). 
\ee
The phase diagram must be symmetric under these transformations. 

\begin{figure}
\centering
\includegraphics[scale=1.]{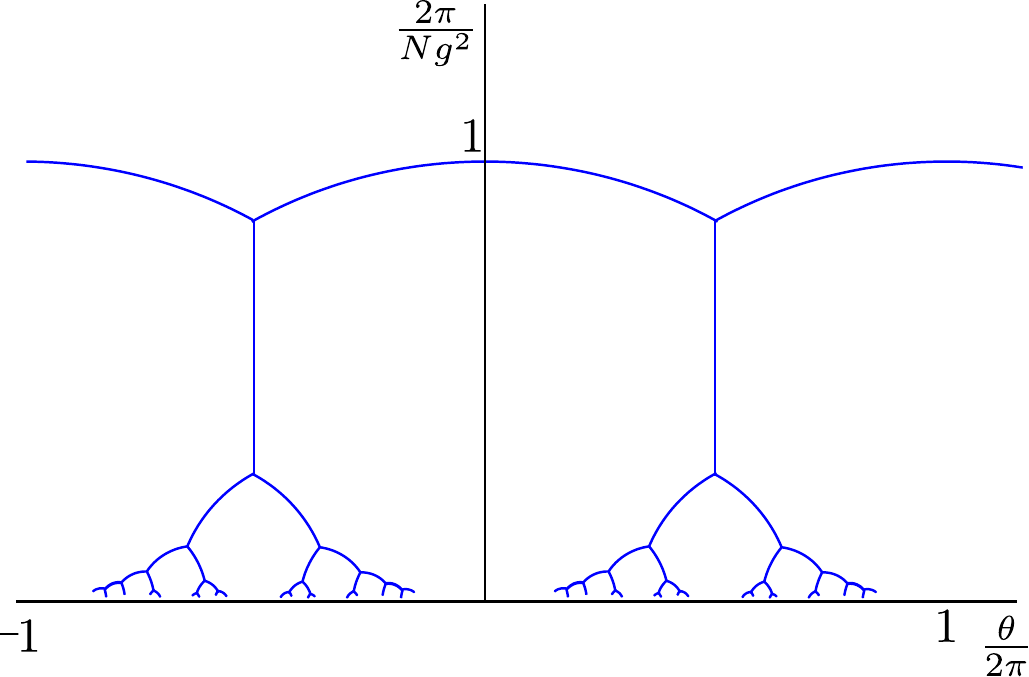}
\caption{The phase diagram of the model in the $\tau$ plane, 
with $\tau=\frac{\theta}{2\pi}+\im \frac{2\pi}{N g^2}$. 
We here draw the case, where the system is gapped at any $\tau$. 
The blue curves denote the phase transitions, and they are related by the self-dual transformations, $SL(2,\mathbb{Z})$. }
\label{fig:phase_diagram2}
\end{figure}

This provides the motivation to draw the phase diagram in the $\tau$ plane~\cite{Cardy:1981fd}, 
and we show it in Fig.~\ref{fig:phase_diagram2}
with the horizontal axis ${\rm Re}(\tau) =\frac{\theta}{2\pi}$ and 
vertical axis ${\rm Im}(\tau) =\frac{2\pi}{Ng^2}$. 
This shows the same phase diagram with the left figure in Fig.~\ref{fig:phase_diagram1}, 
and the blue solid curves show the phase transitions between gapped phases. 
These curves are related by $SL(2,\mathbb{Z})$ transformations, and the phase diagram shown in Fig.~\ref{fig:phase_diagram2} is invariant under $SL(2,\mathbb{Z})$. 
This is obvious by noticing that 
\be
\ve_{n,m} (\tau ,\bar{\tau} )
=\frac{N}{\mathrm{Im}(\tau)} |n+m\tau|^2,  
\ee
and thus the criterion (\ref{eq:ansatz_free_energy}) is $SL(2,\mathbb{Z})$ invariant\footnote{
It may be convenient to note that 
$\ve_{n,m} (\tau ,\bar{\tau} )$ appears in 
the summand of the non-holomorphic Eisenstein series
$\displaystyle E_s (\tau ) =\frac{1}{2} \sum_{(n,m)\in\mathbb{Z}^2\setminus (0,0)} \left(\frac{({\rm Im}\tau )}{|m+n\tau |^{2}}\right)^s $, which is manifestly invariant under $SL(2,\mathbb{Z})$. 
}. 
In Fig.~\ref{fig:phase_diagram2}, the Higgs and the confinement phases are separated by the curve, $|\tau|=1$, for $-\frac{1}{2}<\mathrm{Re}(\tau)<\frac{1}{2}$. 
Assuming the existence of this Higgs-confinement phase transition, then all the other phase transitions are obtained by the duality. 

For instance, 
we can obtain the $\mathsf{CP}$-broken line at $\theta=\pi$ 
from the Higgs-confinement phase transition curve as follows.
$ST^{-1}$ maps $\tau$ and charges as
\begin{\eq}
ST^{-1}:\ \tau \to -\frac{1}{\tau -1} ,\quad
(n,m)\to (-m , n+m) .
\end{\eq}
Therefore,
the Higgs-confinement phase transition curve is mapped as
\begin{\eq}
ST^{-1}:\ \tau =\rme^{\im \phi} \to \frac{1}{2} +\frac{\im}{2}\cot{\frac{\phi}{2}}
\quad \left( \frac{\pi}{3}< \phi <\frac{2\pi}{3} \right) ,
\end{\eq}
while the charges $(1,0)$ and $(0,1)$ are mapped as
\begin{\eq}
ST^{-1}:\ (1,0)\to (0 , 1) ,\quad  (0,1) \to (-1,1 ) .
\end{\eq}
This implies that the phase boundary between condensations of the charges $(1,0)$ and $(0,1)$
is mapped to the one between $(0,1)$ and $(-1,1)$ at $\theta =\pi$,
which is nothing but the CP-broken line. 
Figure~\ref{fig:phase_diagram2} also clarifies the existence of various oblique confinement phases in the strong coupling regime of this model. 
For example, the region with the condensation of charge $(-1,\ell )$
can be obtained by applying $ST^{-\ell}$ to the one of the confinement phase.
When $\mathrm{Im}(\tau)$ approaches to $0$, 
there are infinitely many numbers of different oblique confinement phases as a function of $\theta$~\cite{Cardy:1981qy, Cardy:1981fd}. 

Let us emphasize, however, that the $SL(2,\mathbb{Z})$ duality applies only for the local dynamics when $N>1$, and it does not exist as the self duality of the theory. 
In the weak coupling region, $\mathrm{Im}(\tau)\gg 1$, the system is expected to be in the Higgs phase, which is a gapped phase with deconfined Wilson loops, as it is caused by the condensation of electrically charged particles. 
Under the $S$ transformation, it is mapped to the strong coupling region, $\mathrm{Im}(-1/\tau)\ll 1$. 
The vacuum state is now described by the condensation of magnetic monopoles, so the system is again gapped but the Wilson loops are confined. 
This means that the two systems at $\tau$ and $-1/\tau$ are different as topological orders, and the global nature of the low-energy dynamics is not preserved under the $S$ transformation. 

When $N=1$, the Higgs and confinement phases are the same phase
as we do not have an order parameter to distinguish these phases
due to the lack of the one-form symmetry, 
and the phase diagram can be almost trivial as the phase transition curves disappear~\cite{Fradkin:1978dv, Banks:1979fi}.  
We shall concentrate on the case $N>1$ in the following of this paper.

\section{Anomaly and topological aspects of the phase diagram}\label{sec:anomaly_CRmodel}
In this section, we uncover various topological aspects of the Cardy-Rabinovici model when $N>1$. 
For this purpose, 't~Hooft anomaly matching condition plays an important role. 
In order to compute 't~Hooft anomalies of this model, the continuum formulation is more useful than the lattice formulation, mainly because the $\theta$ term can be treated more easily. 
We first give the continuum reformulation of the model, and then study the anomaly matching condition at $\theta=\pi$. 

\subsection{Formal description of the Cardy-Ravinobici model in the continuum}\label{sec:continuum_CRmodel}
In order to analyze the topological aspect of the model, it is desirable to give the continuum formulation. 
However, it is the $U(1)$ gauge theory coupled to both electrically and magnetically charged particles, and we currently do not have the Lagrangian formulation of such a model.

In this paper, therefore, we give up writing down the continuum theory with manifest locality and unitarity, 
and assume that those properties are ensured by the lattice formulation given in (\ref{eq:Lagrangian1}) or (\ref{eq:Lagrangian2}). 
Motivated by (\ref{eq:Lagrangian2}), we express the configuration of the matter fields using their world lines specified by $\{n_\mu\}$ and $\{m_\mu\}$. 
We express the contribution of the electric charge $1$ by the Wilson loop,
\be
W(\{n_\mu\}) = \exp\left(\im \int_{\ell[n_\mu]} a\right) \equiv \exp\left(\im \int a\wedge \star (n_\mu \diff x^\mu)\right), 
\ee
where $a$ is the $U(1)$ gauge field\footnote{Throughout this paper, we use lower cases, $a,b,\ldots$ for dynamical gauge fields, and upper cases $A,B,\ldots$ for background gauge fields. }, and $\ell[n_\mu]$ is the world line corresponding to $\{n_\mu\}$. 
Similarly, the contribution of the magnetic charge $1$ is expressed by the 't~Hooft loop,
\be
H(\{m_\mu\}),
\ee
which is defined as the defect operator \cite{Kapustin:2005py}: 
for sufficiently small two-sphere $S^2$ linking to the loop $\ell[\{m_\mu\}]$, the gauge field must satisfy the quantization condition, 
\be
\int_{S^2} \diff a=2\pi. 
\ee
Then, the ``path integral'' of the matter fields is obtained as\footnote{
It may be more appropriate to use the formulation in \cite{Strassler:1992zr}
but the difference is irrelevant in our argument.
}
\be
Z_{\mathrm{matter}}[a]=\sum_{\{n_\mu\}, \{m_\mu\}} \exp(-S_{\mathrm{matter}}[\{n_\mu\}, \{m_\mu\}])\,  W^N(\{n_\mu\}) H(\{m_\mu\}). 
\label{eq:continuum_matter}
\ee
The Wilson loop appears only in the $N$-th power, as the electric charge of dynamical particles are quantized in $N$. 
We assume that the weight factor, $S_{\mathrm{matter}}[\{n_\mu\}, \{m_\mu\}]$, can be chosen appropriately, so that it is consistent with the axiom of quantum field theories such as locality and unitarity. 
The full partition function is described by summing up the $U(1)$ gauge field, $a$:
\be
Z
=\int \Diff a \, \exp\left(-\frac{1}{2g^2}\int \diff a\wedge \star \diff a
+\im \frac{N \theta}{8\pi^2}\int \diff a \wedge \diff a\right) 
Z_{\mathrm{matter}}[a], 
\label{eq:continuum}
\ee
where $g$ is the coupling constant and $\theta$ is the vacuum angle.

We note that the $\theta$ term in (\ref{eq:continuum}) seems to have an extra factor $N$ in front, and this is necessary for the $2\pi$ periodicity of $\theta$.  
The index theorem, however, tells that 
\be
\frac{1}{8\pi^2}\int \diff a \wedge \diff a \in \mathbb{Z}, 
\ee
on $4$-dimensional spin manifolds, so it may be wondering 
why we need the extra factor $N$ for the $2\pi$ periodicity 
as we would naively expect the $2\pi/N$ periodicity in the convention of (\ref{eq:continuum}). 
Therefore, we need to explain why the naive $2\pi/N$ periodicity is wrong, and how $\theta$ can be still $2\pi$ periodic. 
The key ingredient is the Witten effect~\cite{Witten:1979ey}. 
If we shift $\theta\to \theta+\Delta \theta $ with $\Delta\theta = \frac{2\pi}{N}$, 
then the magnetic particle acquires the extra electric charge $\frac{N\Delta \theta}{2\pi}=1$. 
In other words,
the pure 't~Hooft line is not mapped to itself:
\be
\Bigl\langle H(\{m_\mu\})\Bigr\rangle_{\theta+(2\pi/N)}=\Bigl\langle H(\{m_\mu\}) W^{-1}(\{m_\mu\})\Bigr\rangle_{\theta}. 
\ee
This implies that
if there were no magnetic excitations represented by 't~Hooft lines, the theory has the naive $2\pi/N$ periodicity, but it can be violated by the presence of such excitations. 
Moreover, since only $W^N$ appears in (\ref{eq:continuum_matter}), there is no way to recover the $2\pi/N$ periodicity. 
Under the transformation $\theta\to \theta+2\pi$, however, 
the Witten effect is realized as $H\to H W^{-N}$, 
and the matter partition function is transformed as
\begin{\eq}
 Z_{\mathrm{matter}}[a] \Bigr|_{\theta \rightarrow \theta +2\pi}
= \sum_{\{n_\mu\}, \{m_\mu\}} \exp(-S_{\mathrm{matter}}[\{n_\mu\}, \{m_\mu\}])\,  
W^N(\{n_\mu -m_\mu \}) H(\{m_\mu\}) .
\end{\eq}
So
we have the $2\pi$ periodicity if 
\be
S_{\mathrm{matter}}[\{n_\mu\}, \{m_\mu\}]=S_{\mathrm{matter}}[\{n_\mu-m_\mu\}, \{m_\mu\}]
\quad ({\rm mod}\ 2\pi \im ). 
\ee
We assume this property for the weight factor, and then we identify
\be
\theta\sim \theta+2\pi. 
\label{eq:theta_periodicity}
\ee
We note that this is a stronger statement than saying that the theory is self-dual under the $T$ transformation, $\theta\to \theta+2\pi$. 
Any local observables, including the line operators, should have the same expectation values at $\theta$ and $\theta+2\pi$, and this is true for the Cardy-Rabinovici model. 

Next let us see response to the charge conjugation and $\mathsf{CP}$ transformations.
The charge conjugation $\mathsf{C}$ acts as 
\be
\mathsf{C}: n_\mu\to -n_\mu,\quad  m_\mu\to -m_\mu,
\ee
so the theory is $\mathsf{C}$ symmetric if 
\be
S_{\mathrm{matter}}[\{n_\mu\}, \{m_\mu\}]=S_{\mathrm{matter}}[\{-n_\mu\}, \{-m_\mu\}]
\quad ({\rm mod}\ 2\pi \im ). 
\ee
The $\mathsf{CP}$ transformation, or the time-reversal transformation, acts only on the electric charge,
\be
\mathsf{CP}: n_\mu\to -n_\mu,\quad m_\mu\to m_\mu,
\ee
so the theory also has the $\mathsf{CP}$ symmetry at $\theta=0$ or $\theta=\pi$ if 
\be
S_{\mathrm{matter}}[\{n_\mu\}, \{m_\mu\}]=S_{\mathrm{matter}}[\{-n_\mu\}, \{m_\mu\}] 
\quad ({\rm mod}\ 2\pi \im ). 
\ee
In the lattice model (\ref{eq:Lagrangian1}) or (\ref{eq:Lagrangian2}), 
the weight factor of the matter fields comes out only of the Coulomb interaction, which corresponds to setting $S_{\mathrm{matter}}=0$. 
It obviously satisfies all of the above requirements. 

So far we have discussed the reformulation of the Cardy-Rabinovici model in the continuum limit. 
The remaining topic is the electromagnetic duality. 
Naively thinking, we can realize the electromagnetic duality by requiring
\be
S_{\mathrm{matter}}[\{n_\mu\}, \{m_\mu\}]=S_{\mathrm{matter}}[\{m_\mu\}, \{n_\mu\}]
\quad ({\rm mod}\ 2\pi \im ), 
\ee
as it exchanges the electric and magnetic fields. 
For studying the local dynamics, such as the free energy density, this condition is indeed sufficient to ensure the duality under $S$ as we noted in Sec.~\ref{sec:review_duality}. 
However, the electromagnetic duality, $S$, as a theory is more subtle, and we will come back to this issue later.  
This theory has the electric $\mathbb{Z}_N$ one-form symmetry, $\mathbb{Z}_N^{[1]}$. 
It acts on the Wilson loop as 
\be
\mathbb{Z}_N^{[1]}: W(\ell)\to \rme^{\frac{2\pi \im}{N}} W(\ell). 
\ee
Since $\langle W^N\rangle =1$ in (\ref{eq:continuum_matter}), we can measure the electric charge of the test particle only in mod $N$. 
Similarly, as $\langle H\rangle =1$, the theory does not have the magnetic one-form symmetry. 

\subsection{'t~Hooft anomaly at $\theta=\pi$ and anomaly matching constraint}
The $\mathsf{CP}$ symmetry of the model exists only at $\theta=0$ or $\theta=\pi$. 
This is because the $\mathsf{CP}$ operation effectively flips the sign of $\theta$, $\theta\to-\theta$, and recalling that $\theta\sim \theta+2\pi$, the only invariant points are at $\theta=0,\pi$. 

There is an important mixed anomaly between $\mathbb{Z}_N^{[1]}$ and the $\mathsf{CP}$ symmetry at $\theta=\pi$. 
This turns out to be the same anomaly or global inconsistency with that of $SU(N)$ YM theory at $\theta=\pi$~\cite{Gaiotto:2017yup, Tanizaki:2017bam} (see Refs.~\cite{Kikuchi:2017pcp, Komargodski:2017dmc, Komargodski:2017smk, Shimizu:2017asf, Gaiotto:2017tne, Tanizaki:2017qhf, Tanizaki:2017mtm,  Yamazaki:2017dra, Guo:2017xex, Aitken:2018kky, Tanizaki:2018wtg, Yonekura:2019vyz, Poppitz:2019fnp, Cordova:2018acb, Karasik:2019bxn, Misumi:2019dwq, Anber:2019nze, Anber:2020gig,
Sulejmanpasic:2020zfs, Furusawa:2020qdz} on related studies). 
To see this, let us gauge $\mathbb{Z}_N^{[1]}$ 
by introducing the $\mathbb{Z}_N$ two-form background gauge field, $B$. 
It is constructed as the $U(1)$ two-form gauge field satisfying 
\be
N B=\diff C,
\ee
where $C$ is an auxiliary $U(1)$ gauge field. 
By this condition, $B$ is restricted to have the quantized flux in the unit of $\frac{2\pi}{N}$:
\begin{\eq}
\oint_S B = \frac{1}{N}\oint_S \diff C \in \frac{2\pi}{N} \mathbb{Z},
\end{\eq}
where $S$ is a closed two-dimensional surface.
Postulating the one-form gauge invariance,
\bea
&&B\to B+\diff \lambda, \,\quad  C\to C+N\lambda,\\
&& a\to a-\lambda, 
\eea
the action of the Maxwell theory becomes 
\be
-S_{\mathrm{Maxwell}}
=\int \left[
-\frac{1}{2g^2} (\diff a+B)\wedge \star (\diff a+ B) 
+ \frac{\im\, N\, \theta}{8\pi^2}(\diff a+B)\wedge (\diff a+B) \right]. 
\ee
The first term is
manifestly $\mathsf{CP}$-invariant, but the last term is not. 
Under the $\mathsf{CP}$ operation, $\theta=\pi$ is mapped to $\theta=-\pi$, i.e. the $\theta$ angle is shifted by $-2\pi$ from $\theta=\pi$. 
Therefore, the change of the action is given by
\bea
-\Delta S
&=& -N\frac{\im\,(2\pi)}{8\pi^2}\int (\diff a+B)\wedge (\diff a+B) \nonumber\\
&=& -\frac{\im}{4\pi}\int (N \diff a\wedge \diff a+ 2 \diff a \wedge N B+N B\wedge B) \nonumber\\
&=& -\frac{\im\, N}{4\pi}\int B\wedge B \quad (\bmod \  2\pi\im). 
\eea
This means that the partition function changes its phase under the $\mathsf{CP}$ transformation at $\theta=\pi$, depending on the background gauge field $B$ for $\mathbb{Z}_N^{[1]}$:
\be
\mathsf{CP}: 
Z[B]\to \exp\left( -\frac{\im\, N}{4\pi}\int B^2\right)Z[B]. 
\label{eq:anomaly_CP}
\ee
Now we should ask whether or not we can cancel this phase factor 
by adding appropriate local counterterms.
If there does not exist such counter terms, then we regard that this is the genuine anomaly.
Since the $\mathbb{Z}_N^{[1]}$ gauge symmetry is unbroken in the above computation, it is sufficient to consider the $\mathbb{Z}_N^{[1]}$-invariant local counterterms, which is given by 
\be
\frac{\im\, Nk}{4\pi}\int B^2. 
\ee
Here, the level $k$ should take values in $\mathbb{Z}$ for the gauge invariance, and we identify $k+N\sim k$. 
We define the partition function including this local counterterm as
\be
Z_k[B]=\exp\left( \frac{\im\, Nk}{4\pi}\int B^2\right)Z[B],
\ee
and then the $\mathsf{CP}$ transformation acts as 
\be
\mathsf{CP}: Z_k[B]\to \exp\left(-\frac{\im\, N(2k+1)}{4\pi}\int B^2\right)Z_k[B]. 
\ee
The anomaly can be eliminated if 
\begin{\eq}
2k+1=0 \quad \left( {\rm mod}\ N \right) . 
\end{\eq}
The anomaly is the genuine one only if such $k$ does not exist. 
When $N$ is even, there is no $k\in\mathbb{Z}$ satisfying $2k+1=0$ mod $N$ as $2k+1$ is an odd number. 
Therefore, we find the genuine 't~Hooft anomaly for even $N$. 

When $N$ is odd, we can choose 
\be
k=\frac{N-1}{2}\quad (\mathrm{at}\,\, \theta=\pi)
\ee 
to eliminate the anomalous phase, so there is no anomaly at $\theta=\pi$ for $\mathsf{CP}$ and $\mathbb{Z}_N^{[1]}$. 
We note, however, that we can do the same consideration for $\theta=0$, and then $k$ should be $0$ mod $N$ for the $\mathsf{CP}$-invariant regularization. 
Therefore, when we extend our consideration from a single theory at a given coupling constant to a family of theories parametrized by the couplings, we find the global inconsistency between $\theta=0$ and $\theta=\pi$~\cite{Gaiotto:2017yup, Tanizaki:2017bam}. 
In the following, let us see how the anomaly or global inconsistency is matched by the phase diagram shown in Fig.~\ref{fig:phase_diagram1}, or in Fig.~\ref{fig:phase_diagram2}. 

\subsubsection{Coulomb phase}
According to the free-energy argument in Sec.~\ref{sec:free_energy}, 
the system can be in Coulomb phase if $N$ is sufficiently large, and it is shown as the gray region in the right panel of Fig.~\ref{fig:phase_diagram1}. 
In the Coulomb phase, the only low-energy excitations are basically given by massless free photons, and the low-energy effective theory is just the Maxwell theory. 

The anomaly (\ref{eq:anomaly_CP}) can be matched by the existence of those massless excitations. 
Indeed, the computation of the anomaly (\ref{eq:anomaly_CP}) is performed with the Maxwell Lagrangian, so the anomaly matching condition is obviously satisfied. 

\subsubsection{Higgs phase}
In the Higgs phase, the system is gapped due to the condensation of electrically charged particles. 
We have seen in the left panel of Fig.~\ref{fig:phase_diagram1} (or Fig.~\ref{fig:phase_diagram2}) that
it is realized in the weak-coupling regime, $\frac{2\pi}{N g^2}> \frac{\sqrt{3}}{2}$. 
We can write the low-energy effective action of the Higgs phase as 
\be
-S_{\mathrm{eff}}
=-\int v^2 |\diff \varphi-N a|^2+\im \frac{N\,\theta}{ 8\pi^2}\int \diff a\wedge \diff a, 
\ee
where $v$ is the characteristic size of the vacuum expectation value, and $\varphi$ is the phase field of the scalar field. 
In this phase, the one-form symmetry is spontaneously broken,
\be
\mathbb{Z}_N^{[1]}\to 1,
\ee
as the Wilson loop obeys the perimeter law. 
This fact can be seen very easily in the above effective description. 
In the low-energy limit, we can take $v\to \infty$ as it has the mass dimension $1$, and then the effective action takes finite values only if 
\be
N a =\diff \varphi.
\ee
Especially, the field strength becomes zero in that limit, $\diff a=0$, and then the expectation value of the Wilson loop, $\langle W(\ell)\rangle$,  does not change as we change the loop $\ell$. 
This is true for any higher charge Wilson loops, which can be formally expressed as 
\be
\langle W^n(\ell) \rangle =1 
\ee
for any $n=0, 1, 2,\ldots, N-1$. Therefore, the Higgs phase is the $\mathbb{Z}_N$ topological order. 

Let us see how the anomaly is matched in the low-energy limit, $v\to \infty$. 
Introducing the background gauge field $B$, then the low-energy effective action becomes 
\be
-S_{\mathrm{eff}}
=-\int v^2 |\diff \varphi-N a-C|^2 +\im\frac{N\, \theta}{8\pi^2}\int (\diff a +B)^2,
\ee
where $NB=\diff C$. 
In order to detect the anomaly, we must  make $B$ nontrivial, 
so that, e.g., $\frac{N}{4\pi^2}\int B^2=\frac{1}{N}$. 
For such $B$, the $U(1)$ gauge field $Na+C$ 
cannot be an exact form globally, that is 
\be
\diff \varphi\not = Na+C 
\ee
for any $\varphi$. 
In the limit $v\to \infty$, the effective action diverges, and then 
\be
Z[B]\to 0
\ee
for such nontrivial $B$. This obviously satisfies the anomaly equation (\ref{eq:anomaly_CP}) as the both sides are zero. 
This is how the anomaly is explicitly reproduced in the low-energy effective theory of the Higgs phase. 

\subsubsection{Confinement phase}
As the coupling constant becomes larger, the charge condensation is taken over by the monopole condensation, and the system is in the confinement phase. 
We have found in the left panel of Fig.~\ref{fig:phase_diagram1}
that the confinement phase appears even when $\theta\simeq \pi$ 
if $\frac{1}{2\sqrt{3}}< \frac{2\pi}{Ng^2}<\frac{\sqrt{3}}{2}$. 

For the Cardy-Rabinovici model, the confinement phase is the topologically trivial phase. 
This is  because all the nontrivial Wilson loops show the area law and
none of the symmetries is broken.
In the low-energy limit, it can be formally expressed as
\be
\lim_{\ell\to \infty}\langle W^n(\ell)\rangle =0, 
\ee
for $n=1,2,\ldots, N-1$ mod $N$. Here, $\lim_{\ell\to \infty}$ indicates the limit of considering the larger and larger loops, and the perimeter part of the expectation value is assumed to be eliminated by the renormalization. 

Although the confinement phase is a topologically trivial gapped phase, 
its ground-state energy shows nontrivial $\theta$ dependence, as we have seen in Sec.~\ref{sec:free_energy}. 
As a consequence, when $\theta=\pi$, there are two vacua, one of which has the monopole condensation with charge $(n,m)=(0,1)$ 
while the another has the dyon condensation with charge $(n,m)=(-1,1)$. 
These two vacua are related by $\mathsf{CP}$ at $\theta=\pi$, so the system shows the spontaneous $\mathsf{CP}$ breaking, 
\be
(\mathbb{Z}_2)_{\mathsf{CP}}\to 1. 
\ee

Let us explicitly see how the anomaly (\ref{eq:anomaly_CP}) is reproduced by those two vacuum states. 
For simplicity, we renormalize the vacuum energy so that the monopole-condensed state at $\theta=\pi$ has the partition function,
\be
Z_{\mathrm{mono.}}[B]=1. 
\label{eq:SPT_monopole}
\ee
In this normalization, let us assume that the dyon-condensed state at $\theta=\pi$ has the nontrivial partition function\footnote{
In principle, we should be able to derive this phase factor of the partition function by assuming that the vacuum is given by the dyon condensation instead of the monopole condensation. The phase factor (\ref{eq:SPT_dyon}), however, is given as the local $4$-dimensional action, so it can be easily affected by the different choice of the UV renormalization. In other words, we have to choose the UV renormalization very carefully, so that it is consistent with the phase factor (\ref{eq:SPT_monopole}) for the monopole condensation. In this paper, we avoid this complication by requiring the consistency with the anomaly matching condition. 
},
\be
Z_{\mathrm{dyon}}[B]
=\exp\left( \im \frac{N}{4\pi}\int B^2\right). 
\label{eq:SPT_dyon}
\ee
We note that the absolute value of these partition functions must be the same at $\theta=\pi$, because they are related by the spontaneously-broken $\mathsf{CP}$ transformation. 
However, their phases do not need to be the same under the existence of background $B$ fields, and we assign a specific phase to the dyon-condensed phase to reproduce the anomaly: 
the dyon-condensed phase is a nontrivial symmetry-protected topological (SPT) state with the $\mathbb{Z}_N^{[1]}$ symmetry. 
The full partition function is given by
\be
Z[B]
=Z_{\mathrm{mono.}}[B]+Z_{\mathrm{dyon}}[B]
=1+\exp\left(\im \frac{N}{4\pi}\int B^2\right). 
\ee
Under the $\mathsf{CP}$ transformation, 
\bea
\mathsf{CP}:Z[B]
&=& 1+\exp\left(\im \frac{N}{4\pi}\int B^2\right)\nonumber\\
&\to&  1+\exp\left(-\im \frac{N}{4\pi}\int B^2\right)
=\exp\left(-\im \frac{N}{4\pi}\int B^2\right)Z[B]. 
\eea
This is nothing but the anomaly relation (\ref{eq:anomaly_CP}). 

What would be the consequence of the phase assignment (\ref{eq:SPT_monopole}) and (\ref{eq:SPT_dyon})? 
As only the overall phases of the partition functions differ, the local dynamics inside the four-dimensional bulk is completely identical. 
However, if we put those phases on the spacetime with the boundary, the boundary dynamics are very different due to the anomaly inflow: The monopole condensation does not cause anything interesting, but the dyon condensation leads to the deconfinement of the Wilson loop on the boundary, which is expected to be described by the $U(1)_{-N}$ Chern-Simons theory~\cite{Gaiotto:2017yup} (see also Refs.~\cite{Acharya:2001dz, Anber:2015kea, Sulejmanpasic:2016uwq, Cox:2019aji}). 
It would be very interesting if we can confirm this dynamics from the Cardy-Rabinovici model. 

\subsubsection{Oblique confinement phase near $\theta=\pi$}
When the coupling constant is sufficiently large, 
the system is in the confinement phase near $\theta=0$. 
For non-zero $\theta$, however, more exotic condensations can occur due to the Witten effect, 
and they are the oblique confinement phase. 
When $\theta=\pi$, the composite of monopole and dyon with the charge $(n,m)=(-1,2)$ starts to condense 
when the coupling is strong enough, $\frac{2\pi}{Ng^2}< \frac{1}{2\sqrt{3}}$. 
Here, let us recall that the electric and magnetic charges $(e,m)$ is labeled by $(n,m)$ as in (\ref{eq:label_charges}): $(e,m)=(N(-1+\theta/\pi),2)$. 

In the oblique confinement phase, it should be obvious by the Debye-screening argument that the line operators with the charge $(Nn,m)=(-N,2)$ obeys the perimeter law
\be
\lim_{\ell\to \infty}\langle W^{-N}(\ell ) H^{2}(\ell)\rangle =1 .
\ee
In addition, its mutually local line operators\footnote{
For charge $(n,m)$, mutually local line operators have charges satisfying $nm'-mn' =0\ {\rm mod}\ N$.
} are in the perimeter law as well:
\be
\lim_{\ell\to \infty}\langle W^{n'} (\ell ) H^{m'}(\ell)\rangle =1 ,\quad
\left( (-N)m' -2n' =0\ {\rm mod}\ N \right) ,
\label{eq:oblique_perimeter}
\ee
while the other line operators obey the area law:
\be
\lim_{\ell\to \infty}\langle W^{n'} (\ell ) H^{m'}(\ell)\rangle =0 ,\quad
\left( (-N)m' -2n' \neq 0\ {\rm mod}\ N \right) .
\ee
These conditions imply that patterns of spontaneous symmetry breaking are different between even and odd $N$.
When $N$ is even, 
\eqref{eq:oblique_perimeter} tells us 
the deconfinement of the charge $N/2$ Wilson line~\cite{tHooft:1981bkw},
\be
\lim_{\ell\to \infty}\langle W^{N/2}(\ell)\rangle =1,
\ee
while the other nontrivial Wilson lines are confined,
\be
\lim_{\ell\to \infty}\langle W^{n}(\ell)\rangle=0, 
\ee
for $n\not= 0, N/2$ mod $N$. 
This can be characterized by the spontaneous breaking pattern
\be
\mathbb{Z}_N^{[1]} \to \mathbb{Z}_{N/2}^{[1]}\,, 
\ee
and the resulting low-energy theory is the $\mathbb{Z}_2$ topological order. 
Because of this unusual nature, the oblique confinement phase may provide a new way to find the topological orders in condensed matter systems~\cite{Ye:2016ase}. 

We can show that this is good enough to match the anomaly (\ref{eq:anomaly_CP}) using the similar discussion for the anomaly matching in the Higgs phase. 
Here, let us take another approach instead. 
Anomaly matching can be satisfied if the symmetry is broken to the anomaly-free subgroup, so it is sufficient to show that $\mathbb{Z}_{N/2}^{[1]}$ and $\mathsf{CP}$ does not have an 't~Hooft anomaly. 
Let $\tilde{B}$ is a $\mathbb{Z}_{N/2}$ two-form gauge field, 
\be
\frac{N}{2}\tilde{B} = \diff \tilde{C},
\ee
and we replace $B$ by $\tilde{B}$ in (\ref{eq:anomaly_CP}),
\be
\mathsf{CP}: Z[\tilde{B}] 
\to Z[\tilde{B}] \exp \left( -\frac{\im\, N}{4\pi}\int \tilde{B}^2\right). 
\ee
It may seem that the symmetry is still anomalous in this expression, but we should consider if this anomalous phase can be canceled by adding the local counter term:
\be
Z_k[\tilde{B}]
= Z[\tilde{B}]  \exp \left( \frac{\im\, (N/2) k}{4\pi}\int \tilde{B}^2\right), 
\ee
with $k=0,1,\ldots, N/2-1 \bmod N/2$. Under $\mathsf{CP}$, we obtain 
\be
\mathsf{CP}: Z_k[\tilde{B}] 
\to Z_k[\tilde{B}] \exp \left( -\frac{\im\, N (k+1)}{4\pi}\int \tilde{B}^2\right). 
\ee
Setting $k=-1$ mod $N/2$, the anomalous phase indeed disappears. 
Therefore, the symmetry is spontaneously broken to the anomaly-free subgroup, and the anomaly matching condition is satisfied.

When $N$ is odd, the oblique confinement phase is a trivial phase. It has the mass gap, and the nontrivial Wilson lines are all confined:
\be
\lim_{\ell\to \infty}\langle W^n(\ell)\rangle =0 
\ee
for $n\not=0$ mod $N$. Furthermore, unlike the confined phase, the oblique confinement phase respects the $\mathsf{CP}$ symmetry at $\theta=\pi$, so the ground state is unique. 
Indeed, we should note that the system does not have the genuine 't~Hooft anomaly when $N$ is odd, so the trivially gapped state at $\theta=\pi$ is allowed. 
Even in this situation, the global inconsistency still puts a nontrivial constraint to the phase diagram~\cite{Tanizaki:2017bam, Kikuchi:2017pcp, Tanizaki:2018xto, Komargodski:2017dmc, Karasik:2019bxn, Cordova:2019jnf, Cordova:2019uob}: 
the trivial gapped states at $\theta=0$ and $\pi$ should be distinguished as the $\mathbb{Z}_N^{[1]}$-symmetric SPT states, and there must be a quantum phase transition separating them. 
We can explicitly see this as follows. 
The partition function of the oblique confinement phase should have the following $B$ dependence,
\be
Z_{\mathrm{oblique}}[B]
\propto \exp\left(-\im \frac{N-1}{2} \frac{N}{4\pi}\int B^2\right),
\ee
in order to reproduce (\ref{eq:anomaly_CP}). 
We note that the monopole- and dyon-condensed phases, (\ref{eq:SPT_monopole}) and (\ref{eq:SPT_dyon}), have the different level of this topological action.  
As the level is quantized due to the gauge invariance, there is no continuous way to interpolate these levels without phase transitions as long as the $\mathbb{Z}_N^{[1]}$ symmetry is respected. 
Therefore, existence of the phase transition curves between the confinement and oblique confinement phases in Fig.~\ref{fig:phase_diagram1} is ensured for odd $N$: even though they are both trivial as intrinsic topological orders, they are different SPT states.

\section{Electromagnetic duality for $N=M^2$ and anomaly constraint}\label{sec:gaugedCRmodel}
As we have seen in Sec.~\ref{sec:review}, 
the free-energy density of the Cardy-Rabinovici model enjoys the $SL(2,\mathbb{Z})$ self-duality, and it plays the important role to constraint the phase diagram. 
We note, however, that it is not the self-duality of the theory. 
One way to see this failure is that the $S$ transformation does not preserve the $1$-form symmetry 
as the electric and magnetic line operators are interchanged. 
The same observation has been made for the $S$ duality of $\mathcal{N}=4$ super YM theory in Ref.~\cite{Aharony:2013hda}: Under the $S$ duality, the gauge group is mapped to its GNO dual gauge group with appropriate choice of discrete $\theta$ parameters, and $1$-form symmetry is not necessarily preserved.
As a result, the global nature of each phase is not preserved under $SL(2,\mathbb{Z})$, 
and thus, for example, a topologically trivial phase is mapped to an intrinsic topological order, or vice versa.  

In this section, we obtain the $SL(2,\mathbb{Z})$ self-dual theory out of the Cardy-Rabinovici model with $N=M^2$ 
by gauging a part of the one-form symmetry: 
\begin{\eq}
\mathbb{Z}_M^{[1]} \subset \mathbb{Z}_N^{[1]}= \mathbb{Z}_{M^2}^{[1]}  .
\end{\eq}
After this operations, the theory has the $SL(2,\mathbb{Z})$-gravity mixed anomaly, 
and we can find the further constraint on the phase diagram, 
and it turns out to explain more details on Fig.~\ref{fig:phase_diagram2}. 

\subsection{The $\mathbb{Z}_M^{[1]}$-gauged Cardy-Rabinovici model for $N=M^2$}
Let us construct the self-dual theory out of the Cardy-Rabinovici model. 
It turns out that we can obtain such a theory when $N$ is a square number, say $N=M^2$, by gauging $\mathbb{Z}_M^{[1]}\subset \mathbb{Z}_N^{[1]}=\mathbb{Z}_{M^2}^{[1]}$. 
In order to gauge the $\mathbb{Z}_M^{[1]}$ symmetry~\cite{Kapustin:2014gua, Aharony:2013hda}, 
we introduce the dynamical $\mathbb{Z}_M$ $2$-form gauge field, 
\be
M b = \diff c,
\ee
where $c$ is a $U(1)$ $1$-form gauge field and $b$ is a $U(1)$ $2$-form gauge field. 
By this condition, the period of $b$ is quantized in the unit of $\frac{2\pi}{M}$.
We require the $1$-form gauge invariance,
\bea
b\to b+\diff \lambda^{(1)}, \quad c\to c+M\lambda^{(1)} ,\quad
a\to a-\lambda^{(1)} . 
\eea
In order to establish the $1$-form gauge invariance, 
we should replace the field strength $\diff a$ by
\be
\diff a + b = \frac{1}{M}(M \diff a + \diff c) .
\ee
Using the $1$-form gauge symmetry, we can set $a=0$, which is an analogue of the unitary gauge in the ordinary gauge symmetry, and then this formula tells the replacement rule of the field strength $\diff a$ by  $\frac{1}{M}\diff c$.  

The continuum formulation, (\ref{eq:continuum_matter}) and (\ref{eq:continuum}), of the Cardy-Rabinovici model uses the line operators $W(\ell)$ and $H(\ell)$ in its definition, 
so we need to discuss how the $\mathbb{Z}_M^{[1]}$ gauging affects 
the line operators in order to define the gauged theory. 
The fundamental Wilson line, $W(\ell)$, is no longer a genuine line operator as it is not gauge invariant under the $1$-form gauge symmetry, so it becomes 
\be
W(\ell,\Sigma)=\exp\left(\im \int_\ell a+\im \int_\Sigma b\right)
\ee
with $\p \Sigma = \ell$. The electric genuine line operator is given by its $M$-th power,
\be
\widetilde{W}(\ell)\equiv W^M(\ell,\Sigma)=\exp\left(\im \int_\ell (M a + c)\right). 
\ee
In the gauge $a=0$, this is nothing but the fundamental Wilson line for the new $U(1)$ gauge field $c$. 
In the path integral (\ref{eq:continuum_matter}), only $W^N$ appears, which is equal to $\widetilde{W}^M$ after gauging, as we have set $N=M^2$. 

Next let us discuss the magnetic line operator. 
The 't~Hooft line $H(\ell)$ is defined as the defect operator specified the magnetic flux around it. 
Since we should replace the field strength, its definition is changed as 
\be
\int_{S^2}\frac{1}{M}(M\diff a + \diff c)=2\pi
\ee
for small two-spheres $S^2$ linking to $\ell$. In the gauge $a=0$, this reads 
\be
\int_{S^2}\diff c=2\pi M. 
\ee
Let us denote the minimal 't~Hooft line for the gauge field $c$ as $\widetilde{H}(\ell)$, then this equality means that  the original 't~Hooft line $H(\ell)$ should be regarded as the $M$-th power of $\widetilde{H}(\ell)$,
\be
H(\ell)=\widetilde{H}^M(\ell). 
\ee
By combining these data, the partition function of the $\mathbb{Z}_M^{[1]}$-gauged theory is given as 
\bea
Z
&=&\int \Diff c \, 
\exp\left( -\frac{1}{2 (N g^2)}\int \diff c\wedge \star \diff c 
  +\im \frac{\theta}{8\pi^2}\int \diff c \wedge \diff c\right)\nonumber\\
&&\times \sum_{\{n_\mu\}, \{m_\mu\}} \exp(-S_{\mathrm{matter}}[\{n_\mu\}, \{m_\mu\}])\,  \widetilde{W}^M(\{n_\mu\}) \widetilde{H}^M(\{m_\mu\}). 
\label{eq:partition_function_gauged}
\eea
This is the $\mathbb{Z}_M^{[1]}$-gauged Cardy-Rabinovici model for $N=M^2$. 

\subsection{Symmetry, self-duality and anomaly}
Let us discuss the topological properties of the $\mathbb{Z}_M^{[1]}$-gauged Cardy-Rabinovici model based on symmetries, dualities, and anomalies. 
We would like to emphasize that some aspects of these analyses have an important implication to the original Cardy-Rabinovici model with $N=M^2$. 
Gauging of the $\mathbb{Z}_M^{[1]}$ symmetry does not affect the local dynamics. For example, non-topological degeneracy of the vacua should be in common for these two theories, so they have the same phase diagram, Fig.~\ref{fig:phase_diagram2}, while the topological characterization of each phases is affected by gauging $\mathbb{Z}_M^{[1]}$. 

\subsubsection{1-form symmetry and its anomaly}
The $1$-form symmetry of the original Cardy-Rabinovici model is $\mathbb{Z}_N^{[1]}=\mathbb{Z}_{M^2}^{[1]}$ acting on the Wilson loop $W$, $W\to \rme^{2\pi\im/M^2}W$. 
By gauging its $\mathbb{Z}_M^{[1]}$ subgroup, we find the $1$-form symmetry,
\be
(\mathbb{Z}_M^{[1]})_{\mathrm{ele.}}\times (\mathbb{Z}_M^{[1]})_{\mathrm{mag.}}, 
\ee
each factor of which acts on $\widetilde{W}$ and $\widetilde{H}$, respectively, i.e. 
\begin{\eq}
(\mathbb{Z}_M^{[1]})_{\mathrm{ele.}}:\widetilde{W}\to 
\rme^{\frac{2\pi\im}{M}}\widetilde{W} ,\quad
(\mathbb{Z}_M^{[1]})_{\mathrm{mag.}}:\widetilde{H}\to
 \rme^{\frac{2\pi\im}{M}}\widetilde{H} . 
\end{\eq}

In the gapped phase of the gauged model, the $1$-form symmetry is always spontaneously broken. 
In the Higgs phase, the electric lines are deconfined,
\be
\lim_{\ell\to \infty}\langle \widetilde{W}^n(\ell)\rangle = 1,
\ee
for $n=1,\ldots, M-1$, while the magnetic lines are confined,
\be
\lim_{\ell\to \infty}\langle \widetilde{H}^n(\ell)\rangle = 0.
\ee 
In the monopole-condensation phase, the opposite is true. The electric lines are confined, while the magnetic lines are deconfined, 
\be
\lim_{\ell\to \infty}\langle \widetilde{W}^n(\ell)\rangle = 0,\quad 
\lim_{\ell\to \infty}\langle \widetilde{H}^n(\ell)\rangle = 1. 
\ee
In another confinement phase, say the dyon-condensation phase with charge $(-1,1)$, the deconfined lines are 
\be
\lim_{\ell\to \infty}\langle (\widetilde{W}^{-1}\widetilde{H})(\ell)\rangle = 1,
\ee
for $n=1,\ldots, M-1$, and other nontrivial lines are all confined. 
In all of these situations, we find the symmetry breaking pattern,
\be
\left( \mathbb{Z}_M^{[1]} \right)_{\mathrm{ele.}}\times 
\left( \mathbb{Z}_M^{[1]} \right)_{\mathrm{mag.}}\to \mathbb{Z}_M^{[1]},
\label{eq:SSB_1form}
\ee
and the unbroken $\mathbb{Z}_M^{[1]}$ symmetry carries the information of the condensation. 

The $U(1)$ pure Maxwell theory enjoys the $U(1)^{[1]}_{\mathrm{ele.}}\times U(1)^{[1]}_{\mathrm{mag.}}$ symmetry, and the $(\mathbb{Z}_M^{[1]})_{\mathrm{ele.}}\times (\mathbb{Z}_M^{[1]})_{\mathrm{mag.}}$ symmetry is its subgroup. 
The $U(1)^{[1]}_{\mathrm{ele.}}\times U(1)^{[1]}_{\mathrm{mag.}}$ symmetry has the mixed anomaly: introducing the background $2$-form gauge fields, $B_{\mathrm{ele.}}$ and $B_{\mathrm{mag.}}$, the theory is no longer gauge invariant in the genuine four dimensions, 
and the gauge invariance requires the anomaly inflow from the $5$-dimensional bulk topological action,
\be
S_{5d}=\frac{\im}{2\pi} \int B_{\mathrm{mag.}}\wedge \diff B_{\mathrm{ele.}}. 
\ee
Even when restricting the symmetry to $(\mathbb{Z}_M^{[1]})_{\mathrm{ele.}}\times (\mathbb{Z}_M^{[1]})_{\mathrm{mag.}}$, this topological action is still nontrivial, and the anomaly matching condition is imposed. 
This anomaly is also necessary in order to reproduce the original $\mathbb{Z}_{M^2}^{[1]}$ symmetry when gauging $(\mathbb{Z}_M^{[1]})_{\mathrm{mag.}}$~\cite{Tachikawa:2017gyf}. 
The spontaneous breaking, (\ref{eq:SSB_1form}), is indeed one of the scenarios matching this 't~Hooft anomaly. 

\subsubsection{$SL(2,\mathbb{Z})$ self duality, $\mathbb{Z}_6$ subgroup, and mixed gravitational anomaly}
The original Cardy-Rabinovici model has the $\mathbb{Z}_{M^2}$ electric $1$-form symmetry but does not have the magnetic $1$-form symmetry. 
When we perform the electromagnetic duality transformation, i.e. the $S$ transformation, these two symmetries should be exchanged, and the theory is mapped to a different theory with the $\mathbb{Z}_{M^2}$ magnetic $1$-form symmetry and without the electric $1$-form symmetry. 
In this sense, the original Cardy-Rabinovici model cannot be a self-dual theory, even though its local dynamics of charges and monopoles shows an interesting self-duality. 

In the $\mathbb{Z}_M^{[1]}$-gauged Cardy-Rabinovici model \eqref{eq:partition_function_gauged}, 
this problem does not exist
since we have same amounts of the electric and magnetic $1$-form symmetries.
Therefore the theory enjoys the $SL(2,\mathbb{Z})$ self-duality, which is the same with that of pure Maxwell theory. 
They are generated by $S$ and $T$ defined in (\ref{eq:S_transformation}) and (\ref{eq:T_transformation}), 
with the complex coupling $\tau=\frac{\theta}{2\pi} +\im \frac{2\pi}{N g^2}$. 
We note that, in this theory, $\theta$ and $\theta+2\pi$ should not be identified as in (\ref{eq:theta_periodicity}). Instead, we can only say that those two points are related by the duality transformation, $T\in SL(2,\mathbb{Z})$, which maps $\tau\to\tau+1$. 
This is because the expectation value of nontrivial 't~Hooft loops do not show the $2\pi$ periodicity,
\be
\langle \widetilde{H}\rangle_{\theta+2\pi}= \langle \widetilde{H}\widetilde{W}^{-1}\rangle_{\theta}\not=\langle\widetilde{H}\rangle_{\theta}. 
\ee
As an identification of $\theta$, the periodicity is extended to 
\begin{\eq}
\theta\sim \theta+2\pi M ,
\end{\eq}
due to the gauging of $\mathbb{Z}_M^{[1]}$, which corresponds to $\tau\sim \tau+M$. 
This should be compared with the similar extension of the $\theta$ periodicity between $SU(N)$ and $SU(N)/\mathbb{Z}_N$ Yang-Mills theories~\cite{Aharony:2013hda}. 

There are some points in the space of $\tau$, which are fixed points under certain subgroups of $SL(2,\mathbb{Z})$. 
There are two important fixed points. The first one is $\tau=\im$, which is the fixed point of 
\be
(\mathbb{Z}_4)_{S}=\{S^k\, |\, k\in \mathbb{Z}\}\subset SL(2,\mathbb{Z}). 
\ee
Also, $\tau=\exp(\pi\im/3) =\frac{1}{2} +\im \frac{\sqrt{3}}{2}$ is the fixed point of 
\be
(\mathbb{Z}_6)_{ST^{-1}}=\{(ST^{-1})^k\, |\, k\in \mathbb{Z}\}\subset SL(2,\mathbb{Z}). 
\ee
Other points with a nontrivial stabilizer subgroup of $PSL(2,\mathbb{Z})$ in the upper half plane can be mapped to either of the above two points by combinations of $S$ and $T$. 
For those fixed points, their stabilizer subgroups of $SL(2,\mathbb{Z})$ are promoted to symmetry from the self-duality of the theory. 

Such a symmetry group may have an 't~Hooft anomaly. 
We here focus on its mixed anomaly with the gravity. 
The key point is that the partition function of the quantum Maxwell theory on generic manifold
is not invariant under $SL(2,\mathbb{Z})$, 
but instead it behaves as the modular form\footnote{
A function $f(\tau ,\bar{\tau})$ is a modular form of weight $(u,v)$
if $f\left( \frac{a\tau +b}{c\tau +d} \right) = (c\tau +d)^u (c\bar{\tau}+d)^v f(\tau ,\bar{\tau} )$.
In the case of $Z_{\mathrm{Maxwell}}(\tau)$, we have $u=\frac{1}{4}(\chi -\sigma)$ and $v=\frac{1}{4}(\chi +\sigma)$.
}~\cite{Witten:1995gf, Verlinde:1995mz}, 
\bea
Z_{\mathrm{Maxwell}}(\tau+1) &=& Z_{\mathrm{Maxwell}}(\tau),\\
Z_{\mathrm{Maxwell}}(-1/\tau) 
&=& \tau^{\frac{1}{4}(\chi-\sigma)} \overline{\tau}^{\frac{1}{4}(\chi+\sigma)}Z_{\mathrm{Maxwell}}(\tau). 
\label{eq:Maxwell_S}
\eea
Here, $\chi$ is the Euler number of the spin four-manifolds, and $\sigma$ is the signature of the spin four-manifolds, which is equal to $\frac{1}{3}$ of the Pontryagin class $p_1$, 
\be
\sigma = \frac{1}{3} \int \frac{1}{8\pi^2}\tr(R\wedge R),
\ee 
and $R$ is the curvature $2$-form. 
On spin four-manifolds, 
the signature $\sigma$ is known to be quantized in $16\mathbb{Z}$ by the Rokhlin's theorem, 
and the generator is the K$3$ surface, $\sigma(\mathrm{K}3)=-16$. 

Using this information, let us compute the mixed anomaly between the subgroup of $SL(2,\mathbb{Z})$ and gravity~\cite{Seiberg:2018ntt}. 
Let us first consider the case $\tau=\im$, where $(\mathbb{Z}_4)_S$ is a symmetry. 
The extra factor under the $S$ transformation in \eqref{eq:Maxwell_S} is 
\be
\tau^{\frac{1}{4}(\chi-\sigma)} \overline{\tau}^{\frac{1}{4}(\chi+\sigma)}
= \rme^{-2\pi \im (\sigma/8)}=1. 
\ee
In the last equality, we use the fact that $\sigma \in 16\mathbb{Z}$ on spin manifolds, and thus the $S$ transformation does not have a mixed 't~Hooft anomaly with gravity. 
We can, however, find an interesting anomaly at 
\be
\tau_*=\exp(\pi\im /3), 
\ee
where $ST^{-1}$ generates the $\mathbb{Z}_6$ symmetry. 
\bea
ST^{-1}:Z_{\mathrm{Maxwell}}(\tau_*)
&\to& (\tau_*-1)^{-\frac{1}{4}(\chi-\sigma)} (\overline{\tau_*}-1)^{-\frac{1}{4}(\chi+\sigma)}
    Z_{\mathrm{Maxwell}}(\tau_*)\nonumber\\
&=&\exp\left(2\pi\im \frac{\sigma}{6}\right)Z_{\mathrm{Maxwell}}(\tau_*). 
\eea
For the K$3$ surface, this anomalous phase is $\exp(-2\pi\im/3)$, 
and thus $(\mathbb{Z}_6)_{ST^{-1}}$ has an order-$3$ mixed anomaly with the signature density. 
This computation is explicitly done for the pure Maxwell theory, 
but the same anomaly should exist for the gauged Cardy-Rabinovici model 
as the anomaly does not change under the continuous $SL(2,\mathbb{Z})$ preserving deformations. 
We note that the following assumption is made to justify this argument: the gauged Cardy-Rabinovici model enjoys the $SL(2,\mathbb{Z})$ self duality and the Lorentz invariance at low-energies. 

Recently, it has been proven that the mixed gravitational anomaly cannot be matched by the topologically ordered phase if the anomaly is detectable on the K$3$ surface~\cite{Cordova:2019jqi, Cordova:2019bsd}. 
As a result, in order to match the anomaly, the system requires certain massless excitations, such as free photons in the Coulomb phase, or the vacuum break the symmetry, 
\be
(\mathbb{Z}_6)_{ST^{-1}}\to (\mathbb{Z}_2)_{\mathsf{C}}. 
\label{eq:Z6_breaking}
\ee
In the left panel of Fig.~\ref{fig:phase_diagram1}, 
there are three degenerate ground states at $\tau=\tau_*$,
where condensations of the three charges $(1,0)$, $(0,1)$ and $(-1,0)$ occur.
connected by the $ST^{-1}$ transformation.
This implies the spontaneous symmetry breaking \eqref{eq:Z6_breaking} of the $\mathbb{Z}_3$ subgroup in $(\mathbb{Z}_6)_{ST^{-1}}$ that matches the anomaly.
Once we have the $\mathbb{Z}_3$ breaking at $\tau=\tau_*$,
we can find similar breaking of $\mathbb{Z}_3$ at other points
by applying $SL(2,\mathbb{Z})$ transformations.
For instance, the phase boundary of condensations of the charges $(0,1)$, $(-1,1)$ and $(-1,2)$ at $\theta =\pi$
can be obtained by applying $ST^{-2}$ to $\tau=\tau_*$.
In the right panel of Fig.~\ref{fig:phase_diagram1}, the system at $\tau_*$ is in the Coulomb phase, and the anomaly matching is again satisfied. 

\section{Summary and discussion}\label{sec:summary}
In this paper, we revisit the phase structure of the four-dimensional lattice $U(1)$ gauge theory with the $\theta$ angle, which was originally proposed by Cardy and Rabinovici \cite{Cardy:1981qy, Cardy:1981fd}. 
This Cardy-Rabinovici model has been expected to show various phase transitions depending on the coupling, $g^2$, and $\theta$ based on heuristic free-energy arguments of possible condensations 
and also on its consistency with the $SL(2,\mathbb{Z})$ self-duality of local dynamics~\cite{Cardy:1981qy, Cardy:1981fd}. 
We show that this model has the mixed anomaly, or global inconsistency, between $\mathbb{Z}_N^{[1]}$ and $\mathsf{CP}$ at $\theta=\pi$ exactly in the same way with $SU(N)$ YM theory. 
We confirm that the proposed phase diagram is consistent with the constraint by the anomaly matching condition. 

In particular, we discuss properties of the oblique confinement phase around $\theta=\pi$ in details. 
This phase is caused by the condensation of composite particles with the charge $(n,m)=(-1,2)$. 
When $N$ is even, the one-form symmetry is spontaneously broken as $\mathbb{Z}_N^{[1]}\to \mathbb{Z}_{N/2}^{[1]}$ and the low-energy physics is described by the $\mathbb{Z}_2$ topological order. 
We show that this is indeed sufficient in order to match the mixed 't~Hooft anomaly for $\mathbb{Z}_N^{[1]}$ and $\mathsf{CP}$ at $\theta=\pi$, so the oblique confinement phase realizes one of the minimal scenarios to match the anomaly. 

For odd $N$, the  genuine anomaly is not present at $\theta=\pi$ and thus the trivially gapped phase is allowed to appear. 
The oblique confinement phase for odd $N$ is indeed such a phase: 
it is gapped because of the condensation and there are no deconfined lines as none of the test particles can have a charge parallel to the charge $(n,m)=(-1,2)$. 
However, the theory has the global inconsistency between $\theta=0$ and $\theta=\pi$, 
and thus the oblique confinement phase at  $\theta=\pi$ for odd $N$ is a different SPT state from the usual confinement phase caused by the monopole condensation. 
These arguments justify the presence of phase transitions, which were conjectured by Cardy and Rabinovici. 

What are the possible implications to the phase diagram of $SU(N)$ YM theory? 
In the 't~Hooft large-$N$ limit, there is a convincing argument showing that the anomaly at $\theta=\pi$ is matched by spontaneous breakdown of $\mathsf{CP}$ symmetry~\cite{Witten:1980sp}, 
and it is supported by the holographic model~\cite{Witten:1998uka} 
and also by semiclassical computations of deformed or softly-broken supersymmetric YM theories~\cite{Davies:2000nw, Shifman:2008ja, Poppitz:2012sw, Poppitz:2012nz, Anber:2013doa, Chen:2020syd}. 
When $N$ is not so large, however, the dynamics at $\theta=\pi$ may be different. 
The presence of anomaly, or global inconsistency, ensures that there must be at least one quantum phase transition as we change $\theta$ from $0$ to $2\pi$~\cite{Gaiotto:2017yup, Tanizaki:2017bam}. 
In Ref.~\cite{Gaiotto:2017yup}, it has been discussed that 
the Coulomb phase may appear in a finite window including $\theta=\pi$ for $N=2$ as one of possible exceptions from the large-$N$ viewpoint. 
Indeed, this possibility can be realized in the Cardy-Rabinovici model as we can see in the right panel of Fig.~\ref{fig:phase_diagram1}. 
We should note, however, that the local dynamics of the YM theory is very different from that of the Cardy-Rabinovici model. 
In the Cardy-Rabinovici model, the Coulomb phase can appear 
if $\frac{C}{N} < \frac{2}{\sqrt{3}}$ according to the free-energy discussion, 
so the Coulomb phase is preferred for larger $N$ instead of smaller ones. 

As another possibility motivated by the Cardy-Rabinovici model, 
there may be a finite window of the oblique confinement phase around $\theta=\pi$ for pure $SU(N)$ YM theories with small $N$. 
As the anomaly involves the one-form symmetry, the anomaly constraint exists even at finite temperatures as the four-dimensional anomaly induces a mixed anomaly between $\mathbb{Z}_N^{[0]}$, $\mathbb{Z}_N^{[1]}$, and $\mathsf{CP}$ for an effective $3$-dimensional theory. 
When compactifying the oblique confinement phase, $\mathbb{Z}_2^{[0]}$ and $\mathbb{Z}_2^{[1]}$ are both spontaneously broken for $N=2$. 
At sufficiently high temperatures, the deconfinement occurs, where $\mathbb{Z}_2^{[0]}$ is spontaneously broken while $\mathbb{Z}_2^{[1]}$ is unbroken. 
In this scenario, the deconfinement temperature must remain nonzero at any values of $\theta$, because the oblique confinement phase cannot be continuously connected to the high-temperature deconfinement phase. 
This statement is true also for $N=3$ if the oblique confinement is realized around $\theta=\pi$ at low temperatures. For odd $N$, the oblique confinement phase does not break any symmetry, while the high-temperature deconfinement phase breaks $\mathbb{Z}_3^{[0]}$. 

In this paper, we have also discussed the $\mathbb{Z}_M^{[1]}$-gauged Cardy-Rabinovici model with the charge $N=M^2$. 
In the original Cardy-Rabinovici model, the $SL(2,\mathbb{Z})$ self-duality found in Ref.~\cite{Cardy:1981fd} is limited to the local aspect of the theory, mainly because the electromagnetic charge lattice is not invariant under $SL(2,\mathbb{Z})$. 
In the gauged model, the $SL(2,\mathbb{Z})$ self-duality is true also for the global aspect of the theory, and we can discuss its anomaly to constrain the phase diagram. 
The gapped phases are always $\mathbb{Z}_M$ topological orders, and the theory also enjoys the $SL(2,\mathbb{Z})$-gravity mixed anomaly. 
Especially at $\tau=\exp(2\pi\im/3)$, the  $(\mathbb{Z}_6)_{ST^{-1}}$ subgroup of $SL(2,\mathbb{Z})$ becomes a symmetry of the theory, and it has the mixed anomaly with the signature density. 
As a consequence of the anomaly matching, the spontaneous breaking $(\mathbb{Z}_6)_{ST^{-1}}\to (\mathbb{Z}_2)_{\mathsf{C}}$ is required.  
It would be interesting 
if one can find a pure anomaly of the duality in our model
which has been studied well for the Maxwell theory in Refs.~\cite{Hsieh:2020jpj,Hsieh:2019iba}. 
It has been known that various four-dimensional theories with $SL(2,\mathbb{Z})$ duality can be constructed out of the two-torus compactification of $6$-dimensional $(2,0)$ theories. 
We also point out that the finite-temperature setup of double-trace deformed YM theory enjoys an emergent Kramers-Wannier duality~\cite{Anber:2015wha}, which is a low-dimensional analogue of duality under the $S$-transformation.
It is quite amusing if certain deformation of such theories can show the rich structure of the phase diagram because of the gravitational-anomaly constraints as studied in this paper.

\acknowledgments
The authors thank Ken Shiozaki, Mithat \"{U}nsal for useful conversations. 
The authors also thank the YITP--RIKEN iTHEMS workshop ``Potential Toolkit to Attack Nonperturbative Aspects of QFT --Resurgence and related topics--'' (YITP-T-20-03) for providing opportunities of useful discussions in completion of this work. 
The work of Y.~T. was partially supported by JSPS KAKENHI Grant-in-Aid for Research Activity Start-up, 20K22350. 

\bibliographystyle{utphys}
\bibliography{./QFT}

\providecommand{\href}[2]{#2}\begingroup\raggedright\begin{thebibliography}{10}

\bibitem{Nambu:1974zg}
Y.~Nambu, ``{Strings, Monopoles and Gauge Fields},''
\href{http://dx.doi.org/10.1103/PhysRevD.10.4262}{{\em Phys. Rev.} {\bfseries
  D10} (1974) 4262}.

\bibitem{Mandelstam:1974pi}
S.~Mandelstam,
  \href{http://dx.doi.org/10.1016/0370-1573(76)90043-0}{``{Vortices and Quark
  Confinement in Nonabelian Gauge Theories},''} in {\em {Phys. Rep. 23 (1976)
  245-249, In *Gervais, J.L. (Ed.), Jacob, M. (Ed.): Non-linear and Collective
  Phenomena In Quantum Physics*, 12-16}}, vol.~23, pp.~245--249.
\newblock
1976.
\newblock

\bibitem{tHooft:1977nqb}
G.~'t~Hooft, ``On the phase transition towards permanent quark confinement,''
  \href{http://dx.doi.org/10.1016/0550-3213(78)90153-0}{{\em Nucl.Phys.B}
  {\bfseries 138} (1978) 1--25}.

\bibitem{Polyakov:1975rs}
A.~M. Polyakov, ``{Compact Gauge Fields and the Infrared Catastrophe},''
\href{http://dx.doi.org/10.1016/0370-2693(75)90162-8}{{\em Phys. Lett.}
  {\bfseries B59} (1975) 82--84}.

\bibitem{tHooft:1981bkw}
G.~'t~Hooft, ``{Topology of the Gauge Condition and New Confinement Phases in
  Nonabelian Gauge Theories},''
\href{http://dx.doi.org/10.1016/0550-3213(81)90442-9}{{\em Nucl. Phys.}
  {\bfseries B190} (1981) 455--478}.

\bibitem{Witten:1979ey}
E.~Witten, ``{Dyons of Charge $e\theta/2\pi$},''
\href{http://dx.doi.org/10.1016/0370-2693(79)90838-4}{{\em Phys. Lett.}
  {\bfseries B86} (1979) 283--287}.

\bibitem{Cardy:1981qy}
J.~L. Cardy and E.~Rabinovici, ``{Phase Structure of Z(p) Models in the
  Presence of a Theta Parameter},''
\href{http://dx.doi.org/10.1016/0550-3213(82)90463-1}{{\em Nucl. Phys.}
  {\bfseries B205} (1982) 1--16}.

\bibitem{Cardy:1981fd}
J.~L. Cardy, ``{Duality and the Theta Parameter in Abelian Lattice Models},''
\href{http://dx.doi.org/10.1016/0550-3213(82)90464-3}{{\em Nucl. Phys.}
  {\bfseries B205} (1982) 17--26}.

\bibitem{Sulejmanpasic:2019ytl}
T.~Sulejmanpasic and C.~Gattringer, ``{Abelian gauge theories on the lattice:
  $\theta$-terms and compact gauge theory with(out) monopoles},''
  \href{http://dx.doi.org/10.1016/j.nuclphysb.2019.114616}{{\em Nucl. Phys.}
  {\bfseries B943} (2019) 114616},
\href{http://arxiv.org/abs/1901.02637}{{\ttfamily arXiv:1901.02637 [hep-lat]}}.

\bibitem{Gattringer:2018dlw}
C.~Gattringer, D.~G\"{o}schl, and T.~Sulejmanpasic, ``{Dual simulation of the
  2d U(1) gauge Higgs model at topological angle $\theta = \pi\,$: Critical
  endpoint behavior},''
  \href{http://dx.doi.org/10.1016/j.nuclphysb.2018.08.017}{{\em Nucl. Phys.}
  {\bfseries B935} (2018) 344--364},
\href{http://arxiv.org/abs/1807.07793}{{\ttfamily arXiv:1807.07793 [hep-lat]}}.

\bibitem{Gaiotto:2017yup}
D.~Gaiotto, A.~Kapustin, Z.~Komargodski, and N.~Seiberg, ``{Theta, Time
  Reversal, and Temperature},''
  \href{http://dx.doi.org/10.1007/JHEP05(2017)091}{{\em JHEP} {\bfseries 05}
  (2017) 091},
\href{http://arxiv.org/abs/1703.00501}{{\ttfamily arXiv:1703.00501 [hep-th]}}.

\bibitem{Gaiotto:2014kfa}
D.~Gaiotto, A.~Kapustin, N.~Seiberg, and B.~Willett, ``{Generalized Global
  Symmetries},'' \href{http://dx.doi.org/10.1007/JHEP02(2015)172}{{\em JHEP}
  {\bfseries 02} (2015) 172},
\href{http://arxiv.org/abs/1412.5148}{{\ttfamily arXiv:1412.5148 [hep-th]}}.

\bibitem{Tanizaki:2017bam}
Y.~Tanizaki and Y.~Kikuchi, ``{Vacuum structure of bifundamental gauge theories
  at finite topological angles},''
  \href{http://dx.doi.org/10.1007/JHEP06(2017)102}{{\em JHEP} {\bfseries 06}
  (2017) 102},
\href{http://arxiv.org/abs/1705.01949}{{\ttfamily arXiv:1705.01949 [hep-th]}}.

\bibitem{Kikuchi:2017pcp}
Y.~Kikuchi and Y.~Tanizaki, ``{Global inconsistency, 't~Hooft anomaly, and
  level crossing in quantum mechanics},''
  \href{http://dx.doi.org/10.1093/ptep/ptx148}{{\em Prog. Theor. Exp. Phys.}
  {\bfseries 2017} (2017) 113B05},
\href{http://arxiv.org/abs/1708.01962}{{\ttfamily arXiv:1708.01962 [hep-th]}}.

\bibitem{Tanizaki:2018xto}
Y.~Tanizaki and T.~Sulejmanpasic, ``{Anomaly and global inconsistency matching:
  $\theta$-angles, $SU(3)/U(1)^2$ nonlinear sigma model, $SU(3)$ chains and its
  generalizations},'' \href{http://dx.doi.org/10.1103/PhysRevB.98.115126}{{\em
  Phys. Rev.} {\bfseries B98} no.~11, (2018) 115126},
\href{http://arxiv.org/abs/1805.11423}{{\ttfamily arXiv:1805.11423
  [cond-mat.str-el]}}.

\bibitem{Komargodski:2017dmc}
Z.~Komargodski, A.~Sharon, R.~Thorngren, and X.~Zhou, ``{Comments on Abelian
  Higgs Models and Persistent Order},''
  \href{http://dx.doi.org/10.21468/SciPostPhys.6.1.003}{{\em SciPost Phys.}
  {\bfseries 6} no.~1, (2019) 003},
\href{http://arxiv.org/abs/1705.04786}{{\ttfamily arXiv:1705.04786 [hep-th]}}.

\bibitem{Karasik:2019bxn}
A.~Karasik and Z.~Komargodski, ``The bi-fundamental gauge theory in 3+1
  dimensions: The vacuum structure and a cascade,''
  \href{http://dx.doi.org/10.1007/JHEP05(2019)144}{{\em JHEP} {\bfseries 05}
  (2019) 144}, \href{http://arxiv.org/abs/1904.09551}{{\ttfamily
  arXiv:1904.09551 [hep-th]}}.

\bibitem{Cordova:2019jnf}
C.~Cordova, D.~S. Freed, H.~T. Lam, and N.~Seiberg, ``{Anomalies in the Space
  of Coupling Constants and Their Dynamical Applications I},''
  \href{http://dx.doi.org/10.21468/SciPostPhys.8.1.001}{{\em SciPost Phys.}
  {\bfseries 8} no.~1, (2020) 001},
  \href{http://arxiv.org/abs/1905.09315}{{\ttfamily arXiv:1905.09315
  [hep-th]}}.

\bibitem{Cordova:2019uob}
C.~Cordova, D.~S. Freed, H.~T. Lam, and N.~Seiberg, ``{Anomalies in the Space
  of Coupling Constants and Their Dynamical Applications II},''
  \href{http://dx.doi.org/10.21468/SciPostPhys.8.1.002}{{\em SciPost Phys.}
  {\bfseries 8} no.~1, (2020) 002},
  \href{http://arxiv.org/abs/1905.13361}{{\ttfamily arXiv:1905.13361
  [hep-th]}}.

\bibitem{Dashen:1970et}
R.~F. Dashen, ``{Some features of chiral symmetry breaking},''
\href{http://dx.doi.org/10.1103/PhysRevD.3.1879}{{\em Phys. Rev.} {\bfseries
  D3} (1971) 1879--1889}.

\bibitem{Witten:1980sp}
E.~Witten, ``{Large N Chiral Dynamics},''
\href{http://dx.doi.org/10.1016/0003-4916(80)90325-5}{{\em Annals Phys.}
  {\bfseries 128} (1980) 363}.

\bibitem{DiVecchia:1980yfw}
P.~Di~Vecchia and G.~Veneziano, ``{Chiral Dynamics in the Large n Limit},''
  \href{http://dx.doi.org/10.1016/0550-3213(80)90370-3}{{\em Nucl. Phys. B}
  {\bfseries 171} (1980) 253--272}.

\bibitem{Ohta:1981ai}
N.~Ohta, ``{Vacuum Structure and Chiral Charge Quantization in the Large $N$
  Limit},'' \href{http://dx.doi.org/10.1143/PTP.66.1408}{{\em Prog. Theor.
  Phys.} {\bfseries 66} (1981) 1408}.
[Erratum: Prog. Theor. Phys.67,993(1982)].

\bibitem{Creutz:1995wf}
M.~Creutz, ``{Quark masses and chiral symmetry},''
  \href{http://dx.doi.org/10.1103/PhysRevD.52.2951}{{\em Phys. Rev.} {\bfseries
  D52} (1995) 2951--2959},
\href{http://arxiv.org/abs/hep-th/9505112}{{\ttfamily arXiv:hep-th/9505112
  [hep-th]}}.

\bibitem{Smilga:1998dh}
A.~V. Smilga, ``{QCD at theta similar to pi},''
  \href{http://dx.doi.org/10.1103/PhysRevD.59.114021}{{\em Phys. Rev.}
  {\bfseries D59} (1999) 114021},
\href{http://arxiv.org/abs/hep-ph/9805214}{{\ttfamily arXiv:hep-ph/9805214
  [hep-ph]}}.

\bibitem{Witten:1998uka}
E.~Witten, ``{Theta dependence in the large N limit of four-dimensional gauge
  theories},'' \href{http://dx.doi.org/10.1103/PhysRevLett.81.2862}{{\em Phys.
  Rev. Lett.} {\bfseries 81} (1998) 2862--2865},
\href{http://arxiv.org/abs/hep-th/9807109}{{\ttfamily arXiv:hep-th/9807109
  [hep-th]}}.

\bibitem{DiVecchia:2017xpu}
P.~Di~Vecchia, G.~Rossi, G.~Veneziano, and S.~Yankielowicz, ``{Spontaneous $CP$
  breaking in QCD and the axion potential: an effective Lagrangian approach},''
  \href{http://dx.doi.org/10.1007/JHEP12(2017)104}{{\em JHEP} {\bfseries 12}
  (2017) 104}, \href{http://arxiv.org/abs/1709.00731}{{\ttfamily
  arXiv:1709.00731 [hep-th]}}.

\bibitem{Aharony:2013hda}
O.~Aharony, N.~Seiberg, and Y.~Tachikawa, ``{Reading between the lines of
  four-dimensional gauge theories},''
  \href{http://dx.doi.org/10.1007/JHEP08(2013)115}{{\em JHEP} {\bfseries 08}
  (2013) 115},
\href{http://arxiv.org/abs/1305.0318}{{\ttfamily arXiv:1305.0318 [hep-th]}}.

\bibitem{Witten:1995gf}
E.~Witten, ``{On S duality in Abelian gauge theory},''
  \href{http://dx.doi.org/10.1007/BF01671570}{{\em Selecta Math.} {\bfseries 1}
  (1995) 383},
\href{http://arxiv.org/abs/hep-th/9505186}{{\ttfamily arXiv:hep-th/9505186
  [hep-th]}}.

\bibitem{Verlinde:1995mz}
E.~P. Verlinde, ``{Global aspects of electric - magnetic duality},''
  \href{http://dx.doi.org/10.1016/0550-3213(95)00431-Q}{{\em Nucl. Phys. B}
  {\bfseries 455} (1995) 211--228},
  \href{http://arxiv.org/abs/hep-th/9506011}{{\ttfamily arXiv:hep-th/9506011}}.

\bibitem{Seiberg:2018ntt}
N.~Seiberg, Y.~Tachikawa, and K.~Yonekura, ``{Anomalies of Duality Groups and
  Extended Conformal Manifolds},''
  \href{http://dx.doi.org/10.1093/ptep/pty069}{{\em PTEP} {\bfseries 2018}
  no.~7, (2018) 073B04}, \href{http://arxiv.org/abs/1803.07366}{{\ttfamily
  arXiv:1803.07366 [hep-th]}}.

\bibitem{Villain:1974ir}
J.~Villain, ``Theory of one- and two-dimensional magnets with an easy
  magnetization plane. ii. the planar, classical, two-dimensional magnet,''
  \href{http://dx.doi.org/10.1051/jphys:01975003606058100}{{\em J. Phys.
  France} {\bfseries 36} no.~6, (1975) 581--590}.

\bibitem{Dirac:1931kp}
P.~A.~M. Dirac, ``{Quantized Singularities in the Electromagnetic Field},''
\href{http://dx.doi.org/10.1098/rspa.1931.0130}{{\em Proc. Roy. Soc. Lond.}
  {\bfseries A133} (1931) 60--72}.

\bibitem{Schwinger:1966nj}
J.~S. Schwinger, ``{Magnetic charge and quantum field theory},''
\href{http://dx.doi.org/10.1103/PhysRev.144.1087}{{\em Phys. Rev.} {\bfseries
  144} (1966) 1087--1093}.

\bibitem{Zwanziger:1968rs}
D.~Zwanziger, ``{Quantum field theory of particles with both electric and
  magnetic charges},''
\href{http://dx.doi.org/10.1103/PhysRev.176.1489}{{\em Phys. Rev.} {\bfseries
  176} (1968) 1489--1495}.

\bibitem{Banks:1977cc}
T.~Banks, R.~Myerson, and J.~B. Kogut, ``{Phase Transitions in Abelian Lattice
  Gauge Theories},'' \href{http://dx.doi.org/10.1016/0550-3213(77)90129-8}{{\em
  Nucl. Phys. B} {\bfseries 129} (1977) 493--510}.

\bibitem{Savit:1977fw}
R.~Savit, ``{Topological Excitations in U(1) Invariant Theories},''
  \href{http://dx.doi.org/10.1103/PhysRevLett.39.55}{{\em Phys. Rev. Lett.}
  {\bfseries 39} (1977) 55}.

\bibitem{Creutz:1979kf}
M.~Creutz, L.~Jacobs, and C.~Rebbi, ``{Experiments with a Gauge Invariant Ising
  System},'' \href{http://dx.doi.org/10.1103/PhysRevLett.42.1390}{{\em Phys.
  Rev. Lett.} {\bfseries 42} (1979) 1390}.

\bibitem{Creutz:1979zg}
M.~Creutz, L.~Jacobs, and C.~Rebbi, ``{Monte Carlo Study of Abelian Lattice
  Gauge Theories},'' \href{http://dx.doi.org/10.1103/PhysRevD.20.1915}{{\em
  Phys. Rev.} {\bfseries D20} (1979) 1915}.

\bibitem{Elitzur:1979uv}
S.~Elitzur, R.~Pearson, and J.~Shigemitsu, ``{The Phase Structure of Discrete
  Abelian Spin and Gauge Systems},''
  \href{http://dx.doi.org/10.1103/PhysRevD.19.3698}{{\em Phys. Rev. D}
  {\bfseries 19} (1979) 3698}.

\bibitem{Horn:1979fy}
D.~Horn, M.~Weinstein, and S.~Yankielowicz, ``{Hamiltonian approach to Z(N)
  lattice gauge theories},''
  \href{http://dx.doi.org/10.1103/PhysRevD.19.3715}{{\em Phys. Rev. D}
  {\bfseries 19} (1979) 3715}.

\bibitem{Fradkin:1978dv}
E.~H. Fradkin and S.~H. Shenker, ``{Phase Diagrams of Lattice Gauge Theories
  with Higgs Fields},'' \href{http://dx.doi.org/10.1103/PhysRevD.19.3682}{{\em
  Phys. Rev. D} {\bfseries 19} (1979) 3682--3697}.

\bibitem{Banks:1979fi}
T.~Banks and E.~Rabinovici, ``{Finite Temperature Behavior of the Lattice
  Abelian Higgs Model},''
  \href{http://dx.doi.org/10.1016/0550-3213(79)90064-6}{{\em Nucl. Phys. B}
  {\bfseries 160} (1979) 349--379}.

\bibitem{Kapustin:2005py}
A.~Kapustin, ``{Wilson-'t Hooft operators in four-dimensional gauge theories
  and S-duality},'' \href{http://dx.doi.org/10.1103/PhysRevD.74.025005}{{\em
  Phys. Rev. D} {\bfseries 74} (2006) 025005},
  \href{http://arxiv.org/abs/hep-th/0501015}{{\ttfamily arXiv:hep-th/0501015}}.

\bibitem{Strassler:1992zr}
M.~J. Strassler, ``{Field theory without Feynman diagrams: One loop effective
  actions},'' \href{http://dx.doi.org/10.1016/0550-3213(92)90098-V}{{\em Nucl.
  Phys. B} {\bfseries 385} (1992) 145--184},
  \href{http://arxiv.org/abs/hep-ph/9205205}{{\ttfamily arXiv:hep-ph/9205205}}.

\bibitem{Komargodski:2017smk}
Z.~Komargodski, T.~Sulejmanpasic, and M.~Unsal, ``{Walls, anomalies, and
  deconfinement in quantum antiferromagnets},''
  \href{http://dx.doi.org/10.1103/PhysRevB.97.054418}{{\em Phys. Rev.}
  {\bfseries B97} no.~5, (2018) 054418},
\href{http://arxiv.org/abs/1706.05731}{{\ttfamily arXiv:1706.05731
  [cond-mat.str-el]}}.

\bibitem{Shimizu:2017asf}
H.~Shimizu and K.~Yonekura, ``{Anomaly constraints on deconfinement and chiral
  phase transition},'' \href{http://dx.doi.org/10.1103/PhysRevD.97.105011}{{\em
  Phys. Rev.} {\bfseries D97} no.~10, (2018) 105011},
\href{http://arxiv.org/abs/1706.06104}{{\ttfamily arXiv:1706.06104 [hep-th]}}.

\bibitem{Gaiotto:2017tne}
D.~Gaiotto, Z.~Komargodski, and N.~Seiberg, ``{Time-reversal breaking in
  QCD$_{4}$, walls, and dualities in 2 + 1 dimensions},''
  \href{http://dx.doi.org/10.1007/JHEP01(2018)110}{{\em JHEP} {\bfseries 01}
  (2018) 110},
\href{http://arxiv.org/abs/1708.06806}{{\ttfamily arXiv:1708.06806 [hep-th]}}.

\bibitem{Tanizaki:2017qhf}
Y.~Tanizaki, T.~Misumi, and N.~Sakai, ``{Circle compactification and 't Hooft
  anomaly},'' \href{http://dx.doi.org/10.1007/JHEP12(2017)056}{{\em JHEP}
  {\bfseries 12} (2017) 056},
\href{http://arxiv.org/abs/1710.08923}{{\ttfamily arXiv:1710.08923 [hep-th]}}.

\bibitem{Tanizaki:2017mtm}
Y.~Tanizaki, Y.~Kikuchi, T.~Misumi, and N.~Sakai, ``{Anomaly matching for phase
  diagram of massless $\mathbb{Z}_N$-QCD},''
  \href{http://dx.doi.org/10.1103/PhysRevD.97.054012}{{\em Phys. Rev.}
  {\bfseries D97} (2018) 054012},
\href{http://arxiv.org/abs/1711.10487}{{\ttfamily arXiv:1711.10487 [hep-th]}}.

\bibitem{Yamazaki:2017dra}
M.~Yamazaki, ``{Relating 't Hooft Anomalies of 4d Pure Yang-Mills and 2d
  $\mathbb{CP}^{N-1}$ Model},''
  \href{http://dx.doi.org/10.1007/JHEP10(2018)172}{{\em JHEP} {\bfseries 10}
  (2018) 172},
\href{http://arxiv.org/abs/1711.04360}{{\ttfamily arXiv:1711.04360 [hep-th]}}.

\bibitem{Guo:2017xex}
M.~Guo, P.~Putrov, and J.~Wang, ``{Time reversal, SU(N) Yang-Mills and
  cobordisms: Interacting topological superconductors/insulators and quantum
  spin liquids in 3+1D},''
  \href{http://dx.doi.org/10.1016/j.aop.2018.04.025}{{\em Annals Phys.}
  {\bfseries 394} (2018) 244--293},
\href{http://arxiv.org/abs/1711.11587}{{\ttfamily arXiv:1711.11587
  [cond-mat.str-el]}}.

\bibitem{Aitken:2018kky}
K.~Aitken, A.~Cherman, and M.~Unsal, ``{Dihedral symmetry in $SU(N)$ Yang-Mills
  theory},''
\href{http://arxiv.org/abs/1804.05845}{{\ttfamily arXiv:1804.05845 [hep-th]}}.

\bibitem{Tanizaki:2018wtg}
Y.~Tanizaki, ``{Anomaly constraint on massless QCD and the role of Skyrmions in
  chiral symmetry breaking},''
  \href{http://dx.doi.org/10.1007/JHEP08(2018)171}{{\em JHEP} {\bfseries 08}
  (2018) 171},
\href{http://arxiv.org/abs/1807.07666}{{\ttfamily arXiv:1807.07666 [hep-th]}}.

\bibitem{Yonekura:2019vyz}
K.~Yonekura, ``{Anomaly matching in QCD thermal phase transition},''
  \href{http://dx.doi.org/10.1007/JHEP05(2019)062}{{\em JHEP} {\bfseries 05}
  (2019) 062},
\href{http://arxiv.org/abs/1901.08188}{{\ttfamily arXiv:1901.08188 [hep-th]}}.

\bibitem{Poppitz:2019fnp}
E.~Poppitz and T.~A. Ryttov, ``{Possible new phase for adjoint QCD},''
  \href{http://dx.doi.org/10.1103/PhysRevD.100.091901}{{\em Phys. Rev. D}
  {\bfseries 100} no.~9, (2019) 091901},
  \href{http://arxiv.org/abs/1904.11640}{{\ttfamily arXiv:1904.11640
  [hep-th]}}.

\bibitem{Cordova:2018acb}
C.~Cordova and T.~T. Dumitrescu, ``{Candidate Phases for SU(2) Adjoint QCD$_4$
  with Two Flavors from $\mathcal{N}=2$ Supersymmetric Yang-Mills Theory},''
\href{http://arxiv.org/abs/1806.09592}{{\ttfamily arXiv:1806.09592 [hep-th]}}.

\bibitem{Misumi:2019dwq}
T.~Misumi, Y.~Tanizaki, and M.~\"Unsal, ``{Fractional $\theta$ angle, 't Hooft
  anomaly, and quantum instantons in charge-$q$ multi-flavor Schwinger
  model},'' \href{http://dx.doi.org/10.1007/JHEP07(2019)018}{{\em JHEP}
  {\bfseries 07} (2019) 018}, \href{http://arxiv.org/abs/1905.05781}{{\ttfamily
  arXiv:1905.05781 [hep-th]}}.

\bibitem{Anber:2019nze}
M.~M. Anber and E.~Poppitz, ``{On the baryon-color-flavor (BCF) anomaly in
  vector-like theories},''
  \href{http://dx.doi.org/10.1007/JHEP11(2019)063}{{\em JHEP} {\bfseries 11}
  (2019) 063}, \href{http://arxiv.org/abs/1909.09027}{{\ttfamily
  arXiv:1909.09027 [hep-th]}}.

\bibitem{Anber:2020gig}
M.~M. Anber and E.~Poppitz, ``{Generalized 't Hooft anomalies on non-spin
  manifolds},'' \href{http://dx.doi.org/10.1007/JHEP04(2020)097}{{\em JHEP}
  {\bfseries 04} (2020) 097}, \href{http://arxiv.org/abs/2002.02037}{{\ttfamily
  arXiv:2002.02037 [hep-th]}}.

\bibitem{Sulejmanpasic:2020zfs}
T.~Sulejmanpasic, Y.~Tanizaki, and M.~\"{U}nsal, ``{Universality between
  vector-like and chiral quiver gauge theories: Anomalies and domain walls},''
  \href{http://dx.doi.org/10.1007/JHEP06(2020)173}{{\em JHEP} {\bfseries 06}
  (2020) 173}, \href{http://arxiv.org/abs/2004.10328}{{\ttfamily
  arXiv:2004.10328 [hep-th]}}.

\bibitem{Furusawa:2020qdz}
T.~Furusawa, Y.~Tanizaki, and E.~Itou, ``{Finite-Density Massless Two-Color QCD
  at Isospin Roberge-Weiss Point and 't Hooft Anomaly},''
  \href{http://dx.doi.org/10.1103/PhysRevResearch.2.033253}{{\em Phys. Rev.
  Res.} {\bfseries 2} (2020) 033253},
  \href{http://arxiv.org/abs/2005.13822}{{\ttfamily arXiv:2005.13822
  [hep-th]}}.

\bibitem{Acharya:2001dz}
B.~S. Acharya and C.~Vafa, ``{On domain walls of N=1 supersymmetric Yang-Mills
  in four-dimensions},'' \href{http://arxiv.org/abs/hep-th/0103011}{{\ttfamily
  arXiv:hep-th/0103011}}.

\bibitem{Anber:2015kea}
M.~M. Anber, E.~Poppitz, and T.~Sulejmanpasic, ``{Strings from domain walls in
  supersymmetric Yang-Mills theory and adjoint QCD},''
  \href{http://dx.doi.org/10.1103/PhysRevD.92.021701}{{\em Phys. Rev.}
  {\bfseries D92} no.~2, (2015) 021701},
\href{http://arxiv.org/abs/1501.06773}{{\ttfamily arXiv:1501.06773 [hep-th]}}.

\bibitem{Sulejmanpasic:2016uwq}
T.~Sulejmanpasic, H.~Shao, A.~Sandvik, and M.~Unsal, ``{Confinement in the
  bulk, deconfinement on the wall: infrared equivalence between compactified
  QCD and quantum magnets},''
  \href{http://dx.doi.org/10.1103/PhysRevLett.119.091601}{{\em Phys. Rev.
  Lett.} {\bfseries 119} no.~9, (2017) 091601},
\href{http://arxiv.org/abs/1608.09011}{{\ttfamily arXiv:1608.09011 [hep-th]}}.

\bibitem{Cox:2019aji}
A.~A. Cox, E.~Poppitz, and S.~S. Wong, ``Domain walls and deconfinement: a
  semiclassical picture of discrete anomaly inflow,''
  \href{http://arxiv.org/abs/1909.10979}{{\ttfamily arXiv:1909.10979
  [hep-th]}}.

\bibitem{Ye:2016ase}
P.~Ye, T.~L. Hughes, J.~Maciejko, and E.~Fradkin, ``{Composite particle theory
  of three-dimensional gapped fermionic phases: Fractional topological
  insulators and charge-loop excitation symmetry},''
  \href{http://dx.doi.org/10.1103/PhysRevB.94.115104}{{\em Phys. Rev. B}
  {\bfseries 94} no.~11, (2016) 115104},
  \href{http://arxiv.org/abs/1603.02696}{{\ttfamily arXiv:1603.02696
  [cond-mat.str-el]}}.

\bibitem{Kapustin:2014gua}
A.~Kapustin and N.~Seiberg, ``{Coupling a QFT to a TQFT and Duality},''
  \href{http://dx.doi.org/10.1007/JHEP04(2014)001}{{\em JHEP} {\bfseries 04}
  (2014) 001},
\href{http://arxiv.org/abs/1401.0740}{{\ttfamily arXiv:1401.0740 [hep-th]}}.

\bibitem{Tachikawa:2017gyf}
Y.~Tachikawa, ``{On gauging finite subgroups},''
  \href{http://dx.doi.org/10.21468/SciPostPhys.8.1.015}{{\em SciPost Phys.}
  {\bfseries 8} (2020) 015},
\href{http://arxiv.org/abs/1712.09542}{{\ttfamily arXiv:1712.09542 [hep-th]}}.

\bibitem{Cordova:2019jqi}
C.~Cordova and K.~Ohmori, ``{Anomaly Constraints on Gapped Phases with Discrete
  Chiral Symmetry},'' \href{http://dx.doi.org/10.1103/PhysRevD.102.025011}{{\em
  Phys. Rev. D} {\bfseries 102} no.~2, (2020) 025011},
  \href{http://arxiv.org/abs/1912.13069}{{\ttfamily arXiv:1912.13069
  [hep-th]}}.

\bibitem{Cordova:2019bsd}
C.~Cordova and K.~Ohmori, ``{Anomaly Obstructions to Symmetry Preserving Gapped
  Phases},''
\href{http://arxiv.org/abs/1910.04962}{{\ttfamily arXiv:1910.04962 [hep-th]}}.

\bibitem{Davies:2000nw}
N.~M. Davies, T.~J. Hollowood, and V.~V. Khoze, ``{Monopoles, affine algebras
  and the gluino condensate},'' \href{http://dx.doi.org/10.1063/1.1586477}{{\em
  J. Math. Phys.} {\bfseries 44} (2003) 3640--3656},
\href{http://arxiv.org/abs/hep-th/0006011}{{\ttfamily arXiv:hep-th/0006011
  [hep-th]}}.

\bibitem{Shifman:2008ja}
M.~Shifman and M.~Unsal, ``{QCD-like Theories on R(3) x S(1): A Smooth Journey
  from Small to Large r(S(1)) with Double-Trace Deformations},''
  \href{http://dx.doi.org/10.1103/PhysRevD.78.065004}{{\em Phys. Rev.}
  {\bfseries D78} (2008) 065004},
\href{http://arxiv.org/abs/0802.1232}{{\ttfamily arXiv:0802.1232 [hep-th]}}.

\bibitem{Poppitz:2012sw}
E.~Poppitz, T.~Sch\"{a}fer, and M.~\"{U}nsal, ``{Continuity, Deconfinement, and
  (Super) Yang-Mills Theory},''
  \href{http://dx.doi.org/10.1007/JHEP10(2012)115}{{\em JHEP} {\bfseries 10}
  (2012) 115},
\href{http://arxiv.org/abs/1205.0290}{{\ttfamily arXiv:1205.0290 [hep-th]}}.

\bibitem{Poppitz:2012nz}
E.~Poppitz, T.~Sch\"{a}fer, and M.~\"{U}nsal, ``{Universal mechanism of
  (semi-classical) deconfinement and theta-dependence for all simple groups},''
  \href{http://dx.doi.org/10.1007/JHEP03(2013)087}{{\em JHEP} {\bfseries 03}
  (2013) 087},
\href{http://arxiv.org/abs/1212.1238}{{\ttfamily arXiv:1212.1238 [hep-th]}}.

\bibitem{Anber:2013doa}
M.~M. Anber, S.~Collier, E.~Poppitz, S.~Strimas-Mackey, and B.~Teeple,
  ``{Deconfinement in $\mathcal{N}=1$ super Yang-Mills theory on $\mathbb{R}^3
  \times \mathbb{S}^1$ via dual-Coulomb gas and "affine" XY-model},''
  \href{http://dx.doi.org/10.1007/JHEP11(2013)142}{{\em JHEP} {\bfseries 11}
  (2013) 142},
\href{http://arxiv.org/abs/1310.3522}{{\ttfamily arXiv:1310.3522 [hep-th]}}.

\bibitem{Chen:2020syd}
S.~Chen, K.~Fukushima, H.~Nishimura, and Y.~Tanizaki, ``{Deconfinement and
  $\mathcal {CP}$ breaking at $\theta=\pi$ in Yang-Mills theories and a novel
  phase for SU(2)},'' \href{http://dx.doi.org/10.1103/PhysRevD.102.034020}{{\em
  Phys. Rev. D} {\bfseries 102} no.~3, (2020) 034020},
  \href{http://arxiv.org/abs/2006.01487}{{\ttfamily arXiv:2006.01487
  [hep-th]}}.

\bibitem{Hsieh:2020jpj}
C.-T. Hsieh, Y.~Tachikawa, and K.~Yonekura, ``{Anomaly inflow and $p$-form
  gauge theories},'' \href{http://arxiv.org/abs/2003.11550}{{\ttfamily
  arXiv:2003.11550 [hep-th]}}.

\bibitem{Hsieh:2019iba}
C.-T. Hsieh, Y.~Tachikawa, and K.~Yonekura, ``{Anomaly of the Electromagnetic
  Duality of Maxwell Theory},''
  \href{http://dx.doi.org/10.1103/PhysRevLett.123.161601}{{\em Phys. Rev.
  Lett.} {\bfseries 123} no.~16, (2019) 161601},
  \href{http://arxiv.org/abs/1905.08943}{{\ttfamily arXiv:1905.08943
  [hep-th]}}.

\bibitem{Anber:2015wha}
M.~M. Anber and E.~Poppitz, ``{On the global structure of deformed Yang-Mills
  theory and QCD(adj) on $ {\mathrm{\mathbb{R}}}^3\times {\mathbb{S}}^1 $},''
  \href{http://dx.doi.org/10.1007/JHEP10(2015)051}{{\em JHEP} {\bfseries 10}
  (2015) 051}, \href{http://arxiv.org/abs/1508.00910}{{\ttfamily
  arXiv:1508.00910 [hep-th]}}.

\end{thebibliography}\endgroup
\end{document}